\documentclass[preprint]{elsarticle}

\usepackage{lineno,hyperref}
\modulolinenumbers[5]

\journal{Journal of signal processing}

\usepackage{amsmath}
\usepackage{hyperref}
\usepackage{amssymb,amsthm,amsfonts,mathrsfs}
\usepackage{graphicx}
\usepackage{dsfont,enumitem,stackrel}
\usepackage{color}
\usepackage{mathtools,amssymb,bm}
\usepackage{amstext}
\usepackage{array}
\usepackage[ruled]{algorithm}
\usepackage{algorithmic}
\usepackage{cases}
\usepackage{pgfplots}
\usepackage{accents}
\usepackage{caption}
\usepackage{float}
\usepackage{standalone}
\usepackage{subcaption}
\usepackage{tikz}
\usetikzlibrary{shapes,arrows}

\def\change{black}
\def\chang{black}
\newcommand{\RN}[1]{%
	\textup{\uppercase\expandafter{\romannumeral#1}}%
}

\theoremstyle{plain}
\newtheorem{thm}{\textbf{Theorem}}
\newtheorem{lem}{\textbf{Lemma}}
\newtheorem{prop}{\textbf{Proposition}}

\theoremstyle{definition}

\theoremstyle{remark}
\newtheorem*{rem}{\bf Remark}

\newcommand*{\rom}[1]{\expandafter\@slowromancap\romannumeral #1@}
\graphicspath{{./Figures/}} 
\begin{document}

\begin{frontmatter}

\title{ Off-the-grid Recovery of Time and Frequency Shifts with Multiple Measurement Vectors}
\author{Maral Safari, Sajad~Daei, Farzan~Haddadi}
\address{School of Electrical Engineering, Iran University of Science $\&$ Technology}

\begin{abstract}
We address the problem of estimating time and frequency shifts of a known waveform in the presence of multiple measurement vectors (MMVs). This problem naturally arises in radar imaging and wireless communications. Specifically, a signal ensemble is observed, where each signal of the ensemble is formed by a superposition of a small number of scaled, time-delayed, and frequency shifted versions of a known waveform sharing the same continuous-valued time and frequency components. The goal is to recover the continuous-valued time-frequency pairs from a small number of observations. In this work, we propose a semidefinite programming which exactly recovers $s$ pairs of time-frequency shifts from $L$ regularly spaced samples per measurement vector under a minimum separation condition between the time-frequency shifts. Moreover, we prove that the number $s$ of time-frequency shifts scales linearly with the number $L$ of samples up to a log-factor. Extensive numerical results are also provided to validate the effectiveness of the proposed method over the single measurement vectors (SMVs) problem and MUSIC. In particular, we find that our approach leads to a relaxed minimum separation condition and reduced number of required samples.
\end{abstract}


\end{frontmatter}


\section{Introduction}
Over the past few years, there has been a growing interest in using super-resolution, a tool for recovering the high-resolution information from low-pass data. This technique is shown to be useful in many applications such as radar imaging \cite{handbook2008mi}, astronomy \cite{puschmann2005super}, communication systems \cite{luo2006low}, geophysics \cite{khaidukov2004diffraction}, microscopy \cite{mccutchen1967superresolution} and also in the direction of arrival (DOA) estimation \cite{sayed2005network,stoica2005spectral}, in which the aim is to estimate the directions of narrow-band sources by an array of sensors.

In this work, we study the problem of using an antenna array to estimate the time delays and Doppler (frequency) shifts of a known waveform. This problem naturally occurs in active radar imaging \cite{activradarimaging,heckel2016super,sayyari} and multi-path channel identification in wireless communications \cite{heckel2016super}. More precisely, in these applications, a known waveform $x(t)$ is transmitted and reflections from moving sources are received at the $R$-element antenna array. Writing in mathematical terms, we observe a signal ensemble
\begin{align}\label{eq.rel1}
y_m(t)=\sum_{j=1}^sb_{jm}x(t-\overline{\tau}_j){\rm e}^{i2\pi \overline{\nu}_j t},~ m=1,..., R,
\end{align}
at the array where $R$ is the number of array elements, $b_{jm}$ is the attenuation factor corresponding to the time-Doppler shifts $(\overline{\tau}_j, \overline{\nu}_j)$, and $s$ is the number of moving sources.
In active radar imaging, estimating delay and Doppler shifts provides valuable information about the location and relative velocity of the targets in the scene. Besides, in wireless communications \cite{heckel2016super}, model \eqref{eq.rel1} represents a scenario where a mobile user is rapidly moving and sends a known training sequence to a base station (BS) for channel estimation and equalization purposes. In case that the communication channel is frequency selective, the signal arrives at the BS with multiple different delays and Doppler shifts. Estimating the delays and Doppler shifts is necessary for BS in order to remove the inter-symbol interference. 

By taking regularly spaced samples of $y(t)$ in model \eqref{eq.rel1} (see Section \eqref{section.systemmodel} for a detailed description), we have a measurement vector at each antenna composed of $L$ samples. Considering a $R$-element antenna array, we encounter multiple measurement vectors (MMVs) by assuming that the delay-Doppler pairs remain fixed at the output of array. It is worth mentioning that the aforementioned method for accessing MMVs is different from what has been considered in the literature (see for example \cite{atomicmmv,yang2016exact,hezave2020demixing}), since, there, MMVs refer to multiple snapshots in the time domain. However, here we assume $R$ measurements are observed via an $R$-element antenna array. It is also possible to have multiple measurements in time (alternatively meant to be multiple snapshots) by choosing the probing signal $x(t)$ to be periodic\footnote{See \cite[Appendix H]{heckel2016super} for a detailed discussion.}. There are, however, a few constraints imposed by practical scenarios: The probing signal $x(t)$ has finite band-width $B$, the received signals at the array are only observed during a finite interval of length $T$, the delay-Doppler pairs has finite support, i.e. $(\overline{\tau}_j, \overline{\nu}_j)\in [-\tfrac{T}{2},\tfrac{T}{2}]\times [-\tfrac{B}{2},\tfrac{B}{2}]$. By the latter assumptions, and since the effective support of the probing signal in the time and frequency domain must be greater than the delay-Doppler shifts, $x(t)$ and $y(t)$ must be both approximately time- and band-limited. Hence, the natural resolution limit, i.e. the accuracy up to which $(\overline{\tau}_j, \overline{\nu}_j)$ can be uniquely resolved, is proportional to $\tfrac{1}{B}$ and $\tfrac{1}{T}$ in the delay and Doppler directions, respectively. This resolution limit can be achieved by using a standard digital matched filter in order to identify the delay-Doppler pairs. In this paper, we show that this resolution limit can be broken by assuming that the delay-Doppler pair $(\overline{\tau}_j, \overline{\nu}_j)$ can take any continuous values in $[-\tfrac{T}{2},\tfrac{T}{2}]\times [-\tfrac{B}{2},\tfrac{B}{2}]$ and is not constrained to be on a predefined domain of grids which is the case in the well-known theory of compressed sensing (CS) \cite{candes2006robust}. Specifically, using the on-grid assumption in CS, $\ell_{1,2}$ minimization can be applied to recover the unknowns $(\overline{\tau}_j, \overline{\nu}_j)$ from MMVs. However, most CS-based methods\footnote{See e.g. \cite{hyder2010direction, daei2019exploiting,daei2019distribution,daei2019living,daei2018sample,daei2019error,daei2018improved}.} including $\ell_{1,2}$ minimization needs incoherence property which does not generally hold when the grids are fine and hence we encounter an unavoidable basis mismatch between on-the-grid and true delay-Doppler pairs. Our goal in this paper is to estimate the continuous delay-Doppler pairs $(\overline{\tau}_j, \overline{\nu}_j), j=1,..., s$ from these MMVs. To achieve this goal, we propose general atomic norm problems (inspired by \cite{chandrasekaran2012convex}) for two-dimensional (2D) super-resolution in the noise-free and noisy cases equipped with MMVs. To the best of the authors' knowledge, super-resolving 2D continuous parameters from MMVs using the concepts of atomic norm minimization has not been addressed before and indeed most of the prior works can be seen as a special case of it. Further, our proposed problems can be viewed as a continuous counterpart of $\ell_{1,2}$ minimization in CS. However, unlike $\ell_{1,2}$ which is designed for recovering one on-grid parameter, our framework, instead, is seeking to recover two continuous off-grid unknowns (i.e. delay-Doppler pairs). We show that our proposed atomic norm problems can be efficiently solved using a semidefinite programming (SDP). While it is possible to directly consider an SDP formulation for the primal atomic norm problems by utilizing the theory of multilevel Toeplitz matrices in \cite[Theorem 1]{yang2016vandermonde}, we provide an SDP relaxation to the dual solution which allows us to efficiently identify the continuous delay-Dopper pairs. Moreover, we theoretically prove that $s$ delay-Doppler pairs can be recovered using our proposed method with high probability if we take $\mathcal{O}(s\log^2(\frac{s}{R}))$ (up to a log factor) noise-free samples per measurement vector and provided that a certain minimum separation condition between the time-frequency shifts is satisfied. Numerical results also demonstrate that our proposed approach leads to improved (relaxed) minimum separation condition compared to the case of single measurement vector (SMV) problem proposed in \cite{heckel2016super} in both noise-free and noisy cases. Besides, we show through precise analysis and simulations that under a fixed minimum separation between the delay-Doppler pairs, the number of required samples for successful and robust recovery decreases.
\subsection{Related Works and Key Differences}
Conventional subspace-based methods such as MUSIC and ESPRIT \cite{music,schmidt1982signal,ottersten1991performance}, assume that the signal amplitudes $b_{jm}$ are uncorrelated and the covariance matrix corresponding to samples of each array element is low-rank. The performance of these approaches are prone to be corrupted against noise and correlations between sources $b_{jm}$.

The theory of super-resolution using convex optimization is first initiated by Candes et al. in \cite{candes2014towards}. They propose a problem for recovering off-grid time-domain spikes from low-pass Fourier measurements in the SMV case. Tang et al. in \cite{tang2013compressed} study super-resolution problem in the framework of CS. They propose an atomic norm minimization where the frequency spikes of a signal are recovered from its partial time-domain samples. Their work can be regarded as the continuous counterpart of $\ell_{1}$ minimization in CS. The difference with \cite{candes2014towards} lies in the fact that they consider only partial random observations rather than full observations in \cite{candes2014towards}.

{\color{\change}
Hyder et al. in \cite{hyder2010direction} use $\ell_{2,0}$ approximation approach called JLZA-DOA which is an extended version of $\ell_0$  function to extract the joint row sparsity inherent in the signal.
This approach enhances DOA resolution. Besides, their non-convex algorithm unlike the subspace based methods such as MUSIC does not need the number of sources in advance.} 

{\color{\chang}Yang et al. in \cite{yang2016exact,yang2018sample} study recovering signals which share the same 1D frequency parameter from MMVs. They propose an atomic norm framework to solve this problem and study the advantage of using MMVs over SMV theoretically. They show that the availability of MMVs results in relaxed minimum separation condition and reduced number of required measurements. The observed signal is scaled samples of the sum of $s$ sinusoids and coincides with model \eqref{eq.rel1} when the delays are ignored or alternatively when we have only frequency shifts. However, most of the proof techniques in \cite{yang2016exact,yang2018sample} can not be applied to model \eqref{eq.rel1}. In particular, our atomic framework and choice of atoms (which is a multiplication of two Dirichlet kernel) are completely different from those in \cite{yang2016exact,yang2018sample}, since model \eqref{eq.rel1} deals with two unknown parameters in two different domains (i.e. time and frequency shifts). Apart from this, 1D methods (e.g. in \cite{candes2014towards,ottersten1991performance,tang2013compressed,yang2016exact}) and even their straightforward extensions to higher dimensions such as \cite{chi2014compressive,valiulahi2019two,xu2014precise} can not be used in our problem. The difficulty of our model lies in the uniqueness guarantee which requires to show the existence of a random vector dual polynomial with high probability. In contrast to \cite{chi2014compressive,xu2014precise,tang2013compressed,razavikia2019reconstruction} where the randomness in the scalar dual polynomial comes from the uniform sampling, the randomness in our vector dual polynomial is introduced by the random probing signal $x(t)$. Therefore, our uniqueness guarantee is involved with different probabilistic analysis and consequently different concentration of measure results.}
%

{\color{\chang}Heckel et al. in \cite{heckel2016super} tackle the problem of identifying time-frequency shifts in radar scenario. The observed signal at receiver is a scaled superposition of time and frequency shifted versions of a known waveform. Thus, their model can be regarded as a special case of our model \eqref{eq.rel1} where only one element is used at the array. Specifically, an atomic norm approach is provided to recover the continuous delay-Doppler pairs\footnote{Throughout, we occasionally use time-frequency shifts instead of delay-Doppler shifts.} using SMV. As opposed to what is done in \cite{heckel2016super}, we benefit from the common atomic sparsity pattern of MMVs at the outputs of the sensor array. Our work can be viewed as an extension of the SMV work \cite{heckel2016super} to the MMV case. However, this nontrivial generalization comes with major mathematical differences, of which we can mention the uniqueness proof of our proposed atomic problem that deals with vector-valued dual polynomials which is much more challenging than the scalar-valued polynomial used in \cite{heckel2016super}. Moreover, instead of Hoeffding's inequality used in \cite{heckel2016super}, we employ a generalized Bernstein inequality \cite[Lemma 4]{yang2018sample} in the uniqueness proof leading to a bound that outperforms the sample complexity bound of \cite{heckel2016super} even in the SMV case.
}

There also exist other works that might be somehow relevant to our work such as \cite{heckel2016mimo} which deals with recovering three continuous parameters in multiple input multiple output (MIMO) scenario and \cite{li2018atomic, atomicmmv, yang2018sample,bayat2020separating} which consider the MMV framework. However, either their model (and consequently their probabilistic analysis) is different from ours or their work does not use the availability of MMVs (dealing with scalar dual polynomial instead of vector dual polynomial).
\subsection{Notations}
Throughout the paper, scalars are denoted by lowercase letters, vectors by lowercase boldface letters, and matrices by uppercase boldface letters. The $k$th element of a vector $\bm{x}$ is denoted by $x(k)$. The absolute value of a scalar, the element-wise absolute value of a vector and the cardinality of a set are shown by $|\cdot|$. The infinity norm is $\|\bm{z}\|_{\infty}=\underset{k}{\max}~|z_{k}|$. In addition, $\|\cdot\|_{1}$, $\|\cdot\|_{2}$ and $\|\cdot\|_F$ are reserved for $\ell_{1}$, $\ell_{2}$ and Frobenius  norms, respectively. We define $\|\bm{A}\|:=\max_{\|\bm{v}\|_2=1}\|\bm{Av}\|_2$ and $\|\bm{A}\|_{2,\infty}$ := $\underset{j}{\max} \|\bm{a}_j\|_{2}$, where $\bm{a}_j$ denotes the j-th row of a matrix $\bm{A} $. The operator $\langle\cdot,\cdot \rangle_{\mathbb{R}}$ stands for the real part of the inner product of two vectors. We use a 2D index for vectors or matrices. Indeed, by $[{\bm{z}}]_{(k,l)}, k,l=-N,...,N$, we mean that ${\bm{z}}=[z_{(-N,-N)},z_{(-N,-N+1)},...,z_{(-N,N)},z_{(-N+1,-N)},...,z_{(N,N)}]$. The operators $\rm{tr}(\cdot)$ and $(\cdot)^{\rm H}$ represent the trace and Hermitian of a matrix, respectively. $x^{*}$ is the conjugate of $x$. To show that $\bm{A}$ is a positive semidefinite matrix, we write $\bm{A}\succeq 0$. $\mathds{E}[\cdot]$ and $\mathds{P}[\cdot]$ denote the expectation and probability of an event, respectively.
$ {{\mathbb{S}}^{R-1}}=\{\bm{\varphi} \in \mathbb{C}^{R \times 1}:{{\left\| \bm{\varphi}  \right\|}_{2}}=1\} $ denote the unit complex or real sphere. Finally, we use numerical constants $c,\tilde{c},c',c_1,c_2, ...$  which take on different values at
different places.
\section{System Model and Recovery via Convex Optimization}\label{section.systemmodel}
As we assumed earlier, $(\bar\tau_j,\bar{\nu}_j)\in[\tfrac{-T}{2},\tfrac{T}{2}]\times[\tfrac{B}{2},\tfrac{B}{2}]$. Based on $2BT$-Theorem \cite{slepian1976bandwidth},\cite{durisi2012sensitivity}, we can take samples of $y_m(t)$ in the interval  $[\tfrac{-T}{2},\tfrac{T}{2}]$ at rate $\tfrac{1}{B}$. So, we totally have $L:=BT$ samples\footnote{Without loss of generality, we assume that $L$ is an odd integer.} of the form ({\color{black} A detailed proof which is adopted from \cite[Appendix A]{heckel2016super} is provided in {\color{\change} Appendix} \ref{proof.equivalence}}):
\begin{align}\label{eq.sampled_y}
&{{y}_{p{m}}}=\sum\limits_{j=1}^{s}{{{b}_{j{m}}}}\sum\limits_{k,l=-N}^{N}{{{{D}}_{_{N}}}}(\tfrac{l}{L}-{{\tau }_{j}}){D}_{N}(\tfrac{k}{L}-{{\nu }_{j}}){{x}_{p-l}}{{e}^{i2\pi (\tfrac{kp}{L})}} \text{  }\nonumber\\
&p=-N,...,N \quad L=2N+1,\quad  m=1,...,R, 
\end{align} 
where 
\begin{align}\label{eq.diric}
{{{D}}_{N}}(t):=\tfrac{1}{L}\sum\limits_{k=-N}^{N}{{{\rm e}^{i2\pi tk}}}
\end{align}
is the Dirichlet kernel. $\tau_j:=\tfrac{\overline{\tau}_j}{T}$ and $\nu_j:=\tfrac{\overline{\nu}_j}{B}$ are the normalized time-frequency shifts, respectively. $x_l$ is the $l$-th sample of the probing signal $x(t)$ and is assumed to be $L$-periodic {\color{black} \cite[Section 2]{heckel2016super}}. It is easy to verify that $(\tau_j,\nu_j)\in[-\tfrac{1}{2},\tfrac{1}{2}]^2$. Due to the periodicity property, without loss of generality, we assume that $(\tau_j,\nu_j)\in[0,1]^2$. Define atoms $\bm{a}\in\mathbb{C}^{L^2\times 1}$ with elements
\begin{align}\label{eq.a_r}
&{{[\bm{a}(\bm{r})]}_{(k,l)}}={{{D}}_{N}}(\tfrac{l}{L}-\tau ){{{D}}_{N}}(\tfrac{k}{L}-\nu ),\quad \bm{{r} }=[\tau,\nu]^{\top},k,l=-N,...,N.
\end{align}
By using the definition of Dirichlet kernel in \eqref{3}, the atoms $\bm{a}\in\mathbb{C}^{L^2\times 1}$ can also be reformulated as {\color{black}\cite[Section 6]{heckel2016super}}:
\begin{align}\label{eq.ar}
\bm{a}(\bm{r})=\bm{F}^{\rm H}\bm{f}(\bm{r}),
\end{align}
where $[\bm{f}(\bm{r})]_{(m,n)}:={\rm e}^{-i2\pi (m\tau+n\nu)}$,
and $\bm{F}^{\rm H}$ is the inverse 2D discrete Fourier transform whose entries are given by
\begin{align}
[\bm{F}^{\rm H}]_{(k,l),(m,n)}:=\tfrac{1}{L^2}{\rm e}^{i2\pi(\tfrac{m k+n l}{L})}.
\end{align}
The relation \eqref{2} can be reformulated in matrix form as
\begin{align}\label{eq.input_output_rel}
\bm {Y}=\bm{G}\bm{X}\in\mathbb{C}^{L\times R},
\end{align}
where 
\begin{align}\label{signal}
&\bm{X}=\sum_{j=1}^{s}\bm{a}(\bm{r}_j)\bm{b}_j^{\rm H}=\sum_{j=1}^{s}c_j\bm{a}(\bm{r}_j)\bm{\varphi}_j^{\rm H}{\color{\change}=:\sum_{j=1}^sc_j\bm{a}(\bm{r}_j,\bm{\varphi}_j)}.
\end{align}
${{\bm{b}}_{j}^{\rm H}}=[{{b}_{j1}},...,{{b}_{jR}}]\in {\mathbb{C}}^{1 \times R}$ is the attenuation vector,
%
${{c}_{j}}={{\left\| {\bm{b}_{j}} \right\|}_{2}}>0$, and ${{\bm{\varphi}}_{j}}=c_{j}^{-1}{\bm{b}_{j}}\in\mathbb{S}^{R-1}$. Here, $\bm{G}\in {{\mathbb{C}}^{L\times {{L}^{2}}}}$ is the Gabor matrix whose elements are given by 
\begin{align}\label{gabor}
[\bm{G}]_{p,(k,l)}:={{x}_{p-l}}{{\rm e}^{i2\pi (\frac{kp}{L})}},\quad k,l,p=-N,...,N.
\end{align}
We observe from \eqref{signal} that $\bm{X}\in \mathbb{C}^{L^{2}\times R} $ is a sparse combination of a few matrix atoms
$\bm{a}(\bm{r}_j,\bm{\varphi}_j), j=1,..., s$ belonging to the atomic set
\begin{align*}
\mathcal{A} :=\{\bm{a(r,\varphi)}:= \bm{a}(\bm{r})\bm{\varphi}^{\rm H}:\quad \bm{ r}\in {{[0,1]}^{2}},\quad \bm{\varphi} \in {{\mathbb{S}}^{R-1}}\}.
\end{align*}
Hence, to extract $\bm{X}\in\mathbb{C}^{L^2\times R}$ from the underdetermined observations $\bm{Y}\in\mathbb{C}^{L\times R}$, inspired by \cite{chandrasekaran2012convex}, we propose the atomic norm minimization
\begin{align}\label{atommin}
\min_{\bm{Z}\in\mathbb{C}^{L^2\times R}}\|\bm{Z}\|_{\mathcal{A}}~\text{  subject to  } \bm {Y}=\bm{G Z},
\end{align}
where the atomic norm $\|\bm{X}\|_{\mathcal{A}}$ is defined as
\begin{align}\label{atomic}
&{{\left\| \bm{X} \right\|}_{\mathcal{A}}}:=\inf \{t>0: \bm{X}\in {t\,
	{\rm conv}(\mathcal{A})}\}=\nonumber\\
&\inf \{\sum\limits_{j}{{{c}_{j}}:\bm{X}}=\sum\limits_{j}{{{c}_{j}}\bm{a}(\bm{r}_j,\bm{\varphi}_j)}, {{c}_{_{j}}}>0,{{\bm{r}}_{j}}\in {{[0,1]}^{2}}\}, 
\end{align}
and ${\rm conv}(\mathcal{A})$ denotes the convex hull of $\mathcal{A}$.
\section{Main Result}
In the following, we state our main result which provides conditions for exact recovery of $\bm{X}$ in \eqref{atommin}.
\begin{thm}\label{thm.main}
	Suppose that the entries of the probing signal $x_l$ , $l=-N,...,N$, are i.i.d. random variables distributed as $\mathcal{N} (0,\tfrac{1}{L})$ where $L=2N+1$. Let $\bm{Y}\in\mathbb{C}^{L\times R}$ be the observed matrix at the $R$-element antenna array as in \eqref{eq.input_output_rel} i.e. 
	\begin{align}
	\bm{Y}=\bm{G X}\in\mathbb{C}^{L\times R},~\bm{X}=\sum\limits_{{{\bm{r}}_{j}}\in \mathcal{S}}{{{c}_{j}}\bm{a}(\bm{r}_j,\bm{\varphi}_j)}
	\end{align}
	with $L>1024$. Here, $\bm{G}$ is the Gabor matrix defined in \eqref{gabor} and $\mathcal{S}$ is the location of delay-Doppler pairs corresponding to $\bm{X}$. Fix $\delta >0$. Assume that ${{\bm{\varphi} }_{j}}, j=1,..., s$ are independent and uniformly distributed on the unit sphere $\mathbb{S}^{R-1}$ with $\mathds{E}[{{\bm{\varphi} }_{j}}]=0$ and that the set of time-frequency shifts $ \mathcal{S} =\{{{\bm{r}}_{1}},{{\bm{r}}_{2}},...,{{\bm{r}}_{s}}\}\subset {{[0,1]}^{2}}$ obeys the minimum separation condition 
	\begin{align}\label{sep}
	&\max(\Big|\tau_j-\tau_{j'}\Big|,\Big|\nu_j-\nu_{j'}\Big|)\ge\tfrac{2.38}{N} ,\forall(\tau_j,\nu_j),(\tau_{j'},\nu_{j'})\in \mathcal{S}\quad  \text{with}\quad j\ne j'.
	\end{align}
	Moreover, assume that
{\color{black}
		\begin{align}\label{eq.cons}
		L \ge cs \max\{\log^2{\tfrac{12sL}{\delta}}
			(1+\tfrac{1}{R}\log(\tfrac{2L}{\delta})),\log^2\tfrac{18s^2}{\delta}(1+\tfrac{1}{R}\log(\tfrac{L}{\delta}))\}
		\end{align}
	}
	where $c$ is a constant. Then, with probability at least $1-\delta $, $\bm{X}$ is the unique solution of \eqref{atommin}.
\end{thm}
Proof. See Appendix \ref{construction.dual.poly}.

	\begin{rem}(Performance of our result when $R=1$)
		In case of SMV i.e. $R=1$, our bound in \eqref{eq.cons} suggests that the unique recovery is possible under a weaker condition than the previous condition $L\ge cs\log^3(\frac{L^6}{s})$ in the SMV case \cite[Theorem 3.1]{heckel2016super}.
	\end{rem}
	\begin{rem}(Advantages of MMV over $R$ independent SMVs)
	 {\color{\change}By the bound \eqref{eq.cons}, we observe that MMV requires less samples than SMV when $R\ge 2$. Now consider the case where we aim to recover each column of $\bm{X}\in\mathbb{C}^{L^2\times R}$ separately via the SMV method in \cite[Theorem 3.1]{heckel2016super}. When the number of required samples satisfies \eqref{eq.cons} with $R=1$, each column of $\bm{X}$ can be recovered with probability at least $1-\delta$. Hence, $\bm{X}$ can be recovered from $\bm{Y}$ using $R$ independent SMVs with probability at least $1-R\delta$ provided that
		\begin{align}\label{eq.cons1}
		L \ge cs \max\{\log^2{\tfrac{12sL}{\delta}}
	(1+\log(\tfrac{2L}{\delta})),\log^2\tfrac{18s^2}{\delta}(1+\log(\tfrac{L}{\delta}))\}.
		\end{align}
	}
		In contrast, using our proposed MMV problem \eqref{atommin}, one can estimate $s$ time-frequency components with probability $1-R\delta$ provided that
		{\color{\change}
		\begin{align}\label{eq.cons2}
		L \ge cs \max\{\log^2{\tfrac{12sL}{R\delta}}
	(1+\tfrac{1}{R}\log(\tfrac{2L}{R\delta})),\log^2\tfrac{18s^2}{R\delta}(1+\tfrac{1}{R}\log(\tfrac{L}{R\delta}))\}.
		\end{align}
		which is weaker than the condition of $R$ SMVs, i.e. \eqref{eq.cons1}, when $R$ increases. As a result, our proposed MMV needs less number of samples than $R$ independent SMVs.}
	\end{rem}
	\begin{rem}
		Gaussian distribution for the probing signal $x_l,~l=-N,..., N$ is not necessary for our theory to hold and Theorem \ref{thm.main} continues to hold for sub-Gaussian distribution as well. The condition $L\ge 1024$ is more like a technical requirement but not an obstacle in practice as evidenced by our numerical results in Section \ref{section.simulation} which show that the identification of delay-Doppler pairs is also possible when the samples are much less than $1024$. The vectors $\bm{b}_j=[b_{j1},...,b_{jR}]^{T}\in\mathbb{C}^{R\times 1}$ in the radar model \eqref{eq.rel1} describe the attenuation factors corresponding to different antennas and in wireless communications and radar they are assumed to have complex Gaussian distribution \cite{bello1963characterization}. Thus, it is natural to assume that the phases of $\bm{b}_j$s, i.e. $\bm{\phi}_j$s, are independent from each other. The separation between the shifts is necessary for exact and stable recovery and has appeared in all super-resolution theories (such as \cite{tang2013compressed,candes2014towards,valiulahi2019two,heckel2016super,yang2016exact}). However, we should highlight that our separation condition \eqref{sep} is not necessary and a less conservative condition would seem to be enough as evidenced by our simulation results in Section \ref{section.simulation}.	
	\end{rem}
The proof of Theorem \ref{thm.main} is built upon constructing a certain dual certificate for \eqref{atommin}. In the following proposition whose proof is provided in Appendix \ref{proof.propositon_optimal}, we describe the desired form of a valid vector-valued dual certificate which guaranties the optimality of $\bm{X}$ in \eqref{atommin}.
\begin{prop}\label{prop.optimality}
	Assume that $\bm {Y}=\bm{G}\bm{X}$ with $\bm{X}=\sum_{\bm{r}_j\in\mathcal{S}}c_j\bm{a}(\bm{r}_j,\bm{\varphi}_j)$ where $\mathcal{S}$ is the location of delay-Doppler pairs corresponding to $\bm{X}$.
	If there exists a vector-valued dual polynomial $\bm{q}: [0,1]^2\rightarrow \mathbb{C}^{R\times 1}$,
	\begin{align}\label{eq.dual_pol_vector}
	\bm{q}(\bm{r})=\bm{\Lambda}^{\rm H}\bm{G}\bm{a}(\bm{r})
	\end{align}
	satisfying
	\begin{align}\label{dual}
	\begin{array}{cc}
	\bm{q}(\bm{r}_j)=\bm{\varphi}_j,&\bm{r}_j\in\mathcal{S},\\
	\|\bm{q}(\bm{r})\|_2\le 1,& \bm{r}\in[0,1]^2\setminus \mathcal{S},
	\end{array}
	\end{align}
	then $\bm{X}$ is the optimal solution of \eqref{atommin}.
	%
\end{prop} 
The problem \eqref{atommin} involves finding infinitely many variables and can not be directly solved. To deal with this problem, one way is to use multilevel Toeplitz matrices. To practically solve \eqref{atommin}, we first obtain its dual formulation obtained from a standard Lagrangian approach (e.g. see \cite[Chapter 6]{bertsekas2003convex}):
\begin{align}\label{eq.dual_prob}
\max_{\bm{\Lambda}\in\mathbb{C}^{L\times R}}\,{\rm  Re}~ {{\langle \text{ }\!\!\bm{\Lambda},\bm{Y} \rangle }_{{F}}}\text{  subject to   }\left\| {{\bm{G}}^{\rm H}}\bm{\Lambda}\right\|_{\mathcal{A}}^{d}\le 1,
\end{align} 
where $\bm{\Lambda}=[{{\bm{\Lambda}}_{pm}}]\in {{\mathbb{C}}^{L\times R}}, m=1,..., R,  p=-N,..., N$ and 
\begin{align}
\|\bm{V}\|_{\mathcal{A}}^d:=\sup_{\|\bm{Z}\|_{\mathcal{A}}\le 1} {\rm Re}\langle \bm{V}, \bm{Z} \rangle
\end{align}
is the dual norm. Hence, using \eqref{eq.ar}, we have
\begin{align}\label{eq.dualnorm_def}
&\|\bm{G}^{\rm H}\bm{\Lambda}\|_{\mathcal{A}}^d=\sup_{\substack{\|\bm{\varphi}\|_2=1\\\bm{r}\in[0,1]^2}}\langle\bm{G}^{\rm H}\bm{\Lambda}, \bm{a}(\bm{r})\bm{\varphi}^{\rm H}\rangle_F=\nonumber\\
&\sup_{\substack{\|\bm{\varphi}\|_2=1\\\bm{r}\in[0,1]^2}}\langle \bm{\varphi}, (\bm{F}\bm{G}^{\rm H}\bm{\Lambda})^{\rm H}\bm{f}(\bm{r})\rangle=\sup_{\bm{r}\in[0,1]^2}\|(\bm{F}\bm{G}^{\rm H}\bm{\Lambda})^{\rm H}\bm{f}(\bm{r})\|_2,
\end{align}
where we used Holder inequality in the last step. Hence, the constraint of \eqref{eq.dual_prob} becomes equivalent to
\begin{align}\label{eq.constraint1}
&\|(\bm{F}\bm{G}^{\rm H}\bm{\Lambda})^{\rm H}\bm{f}(\bm{r})\|_2^2=\sum_{m=1}^R\bigg|\sum_{k,l=-N}^N[\bm{F}\bm{G}^{\rm H}\bm{\Lambda}]_{(k,l),m}{\rm e}^{i2\pi(k\tau+l\nu)}\bigg|^2\le 1,~\forall \bm{r}\in [0,1]^2.
\end{align}
By replacing \eqref{eq.constraint1}, the dual problem \eqref{eq.dual_prob} involves infinitely many constraints. The following proposition which is an adaptation of \cite[Proposition 2.4]{fernandez2016super} and \cite[Corollary 4.27]{dumitrescu2017positive} provides a tractable sufficient condition for the constraint \eqref{eq.constraint1}.
\begin{prop}\label{prop.boundedlemma}
	Let $\bm{P}=[P_{(k,l),m}]$ be a matrix in $\in\mathbb{C}^{L^2\times R}$ with $k,l=-N,..., N,~ m=1,..., R, L=2N+1$.
	If 
	\begin{align*}
	&\sum_{m=1}^R|\sum_{k,l=-N}^N P_{(k,l),m}{\rm e}^{i2\pi(k\tau+l\nu)}|^2\le 1,~~\forall r \in [0,1]^2,
	\end{align*}
	then there exists a Hermitian positive semidefinite matrix $\bm{Q}\in \mathbb{C}^{L^2\times L^2}$ obeying
	\begin{align}
	&\begin{bmatrix}
	\bm{Q}& \bm{P}\\
	\bm{P}^{\rm H}&\bm{I}_{R}
	\end{bmatrix}\succeq 0,
	{\rm trace}((\bm{\Theta}_k\otimes \bm{\Theta}_l)\bm{Q})=\delta_{(k,l)},~\forall k,l=-N, ..., N,
	\end{align}  
	where
	\begin{align*}
	\delta_{(k,l)}:=\left\{\begin{array}{cc}
	1, & (k,l)=(0,0),\\
	0, & {\rm o.w.}
	\end{array}\right\}
	\end{align*}
	is the indicator function and $\bm{\Theta }_{k}$ stands for the Toeplitz matrix composed of ones on the $k$-th diagonal and zeros elsewhere.
\end{prop}
By exploiting Proposition \ref{prop.boundedlemma}, the dual problem \eqref{eq.dual_prob} is relaxed to the following SDP:
\begin{align}\label{eq.sdp}
&\max_{\substack{\bm{\Lambda}\in\mathbb{C}^{L\times R}\\\bm{Q}\in\mathbb{C}^{L^2\times L^2}, \bm{Q}\succeq 0}}\,{\rm Re}{{\left\langle \text{ }\!\!\bm{\Lambda},\bm{Y} \right\rangle }_{{F}}}\text{  subject to   }\nonumber\\
&\begin{bmatrix}
\bm{Q}& \bm{F}\bm{G}^{\rm H}\bm{\Lambda}\\
\bm{\Lambda}^{\rm H}\bm{G}\bm{F}^{\rm H}&\bm{I}_{R}
\end{bmatrix}\succeq 0,
{\rm trace}((\bm{\Theta}_k\otimes \bm{\Theta}_l)\bm{Q})=\delta_{(k,l)},\nonumber\\
&\forall k,l=-N, ..., N,
\end{align}
\begin{rem}
	The problem \eqref{eq.sdp} is only a relaxation for \eqref{eq.dual_prob}. In fact, the size of $\bm{Q}$ could be larger than $L^2\times L^2$, since the sum of squares expression of a bivariate positive trigonometric polynomial with degree $(L,L)$ might have factors with degree greater than the minimum degree $(L,L)$. However, as simulation results of \cite{heckel2016super,dumitrescu2017positive} indicate, relaxations of minimal degree often lead to optimal solutions in practice\footnote{See \cite[Section 6]{heckel2016super} and \cite[Remark 3.6]{dumitrescu2017positive} for a detailed discussion.}. It is worth noting that the exact SDP formulation for multivariate atomic norm problems such as \eqref{atommin} relies on a Vandermonde decomposition of multilevel Toeplitz matrices investigated in \cite[Theorem 1]{yang2016vandermonde} which needs a checking mechanism involving a rank constraint on the optimal Teoplitz matrix. This issue prohibits an exact SDP characterization for the primal 2-D problem similar to what was done in 1-D problems \cite[Proposition 2.1]{tang2013compressed}.
\end{rem}
When the measurements are contaminated with noise, i.e. $\bm{Y}=\bm{G}\bm{X}+\bm{W}$, we solve the following problem:
\begin{align}
\underset{{\bm{Z}}}{\min} ~\|{\bm{Z}}\|_\mathcal{A} \quad {\rm subject ~ to}\text{       }\|\mathbf{Y}-\bm{G}{\bm{Z}}\|_{F}\le \eta,
\end{align}
where $\eta$ is {\color{\change} an upper-bound on $\|\bm{W}\|_F$}. Moreover, the SDP problem in this case takes the form
\begin{align}\label{sdp_noisy}
&\max_{\substack{\bm{\Lambda}\in\mathbb{C}^{L\times R}\\\bm{Q}\in\mathbb{C}^{L^2\times L^2}, \bm{Q}\succeq 0}}{\rm Re}~\langle\bm{\bm{\Lambda}},\mathbf{Y} \rangle_{F}-\eta\|\bm{\bm{\Lambda}}\|_F\quad  {\rm subject ~to} \nonumber\\
&\begin{bmatrix}
\bm{Q}& \bm{F}\bm{G}^{\rm H}\bm{\Lambda}\\
\bm{\Lambda}^{\rm H}\bm{G}\bm{F}^{\rm H}&\bm{I}_{R}
\end{bmatrix}\succeq 0,
{\rm tr}((\bm{\Theta}_k\otimes \bm{\Theta}_l)\bm{Q})=\delta_{(k,l)},\nonumber\\
&\forall k,l=-N, ..., N,
\end{align}
Now, we are ready to state the procedure of finding delay-Doppler pairs from the dual solution in both noiseless and noisy cases. Write the vector-valued dual polynomial
\begin{align}\label{eq.dual_pol}
\bm{q}(\bm{r})=\widehat{\bm{\Lambda}}^{\rm H}\bm{G}\bm{a}(\bm{r}),
\end{align}
where $\widehat{\bm{\Lambda}}$ is the solution to the SDP problems \eqref{eq.sdp} and \eqref{sdp_noisy}. 
Proposition \ref{prop.optimality} suggests that an estimate $\widehat{\mathcal{S}}$ of the delay-Doppler pairs $\mathcal{S}$ can be obtained from
\begin{align}\label{eq.support_estimates}
\widehat{\mathcal{S}}=\{\bm{r}\in [0,1]^2\big|\|\bm{q}(\bm{r})\|_2=1\}.
\end{align}
One way to identify the support $\widehat{\mathcal{S}}$ is by discretizing $\bm{r}\in [0,1]^2$ on a fine domain of grids. Then, one can check the locations $\bm{r}$ that $\|\bm{q}(\bm{r})\|_2$ achieves to one on the grid. We use this heuristic in Section \ref{section.simulation}.
\section{Simulation Results}\label{section.simulation}
\begin{figure}[t]
	
	\begin{subfigure}[b]{.5\textwidth}
		\centering
		\includegraphics[scale=.2]{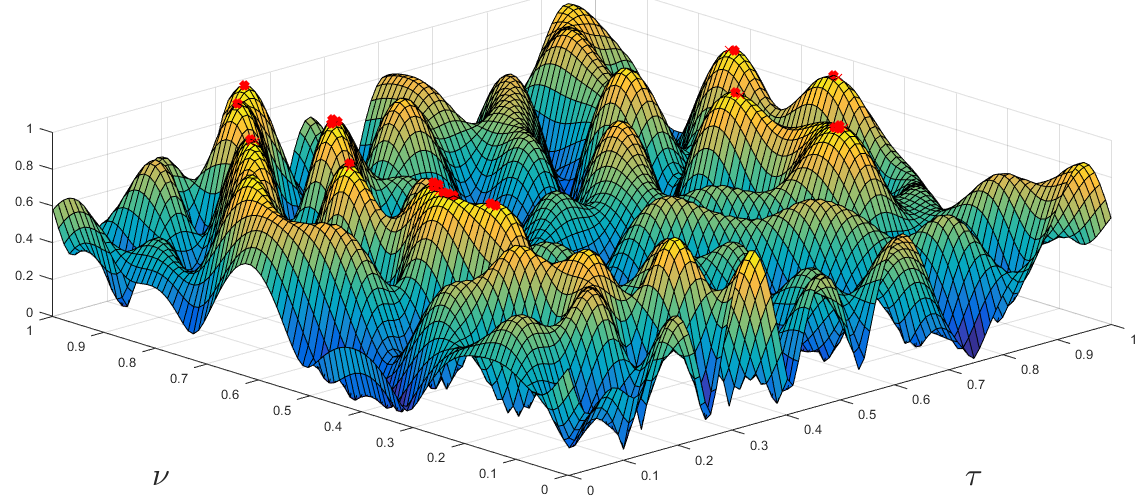}
	\end{subfigure}
	\begin{subfigure}[b]{.5\textwidth}
		\centering
		\includegraphics[scale=.2]{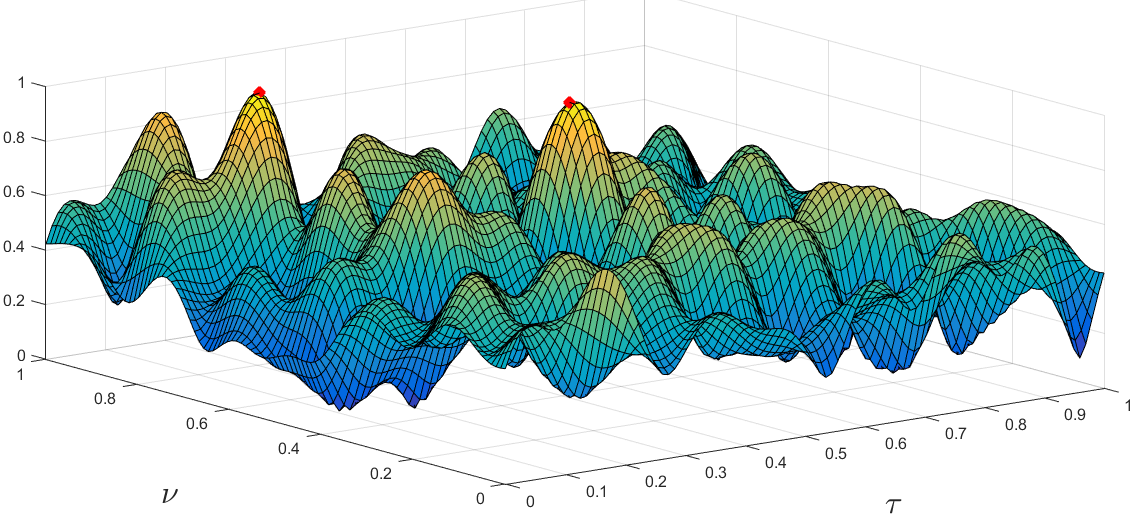}
	\end{subfigure}

	\caption{{\color{\change}Noiseless case. The true sources are located at $\bm{r}_1=(0.2,0.8)$, $\bm{r}_2=(0.5,0.5)$. We set $N=4$ and solve \eqref{eq.sdp}. Left and Right images show $\|\bm{q}(\bm{r})\|_2$ for SMV ($R=1$) and MMV ($R=10$) cases, respectively. Red markers show where $\|\bm{q}(\bm{r})\|_2$ equals one.}}	
	\label{fig1}
\end{figure}

\begin{figure}[t]
	
	\begin{subfigure}[b]{.5\textwidth}
		\centering
		\includegraphics[scale=.2]{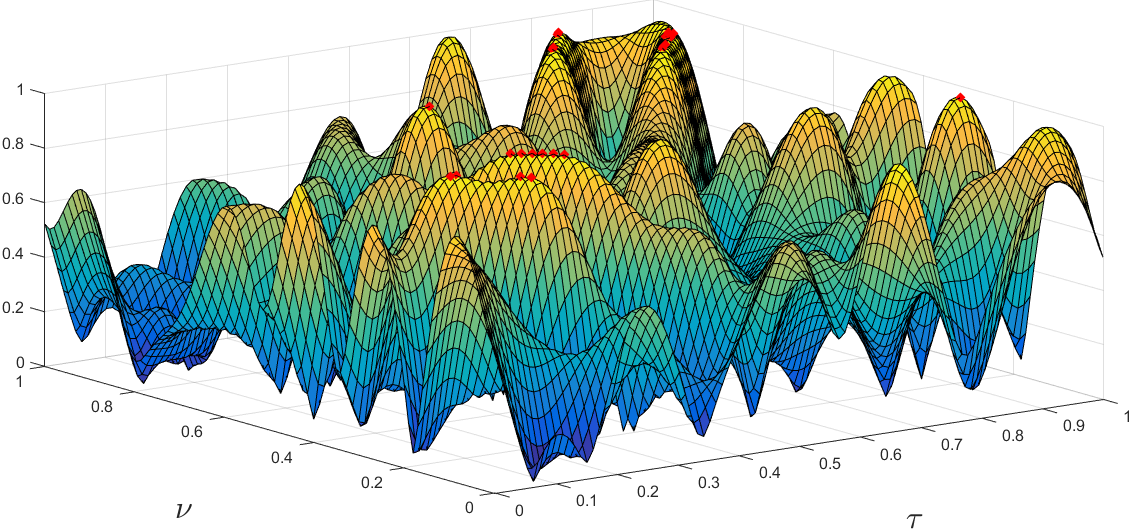}
	\end{subfigure}
	\begin{subfigure}[b]{.5\textwidth}
		\centering
		\includegraphics[scale=.2]{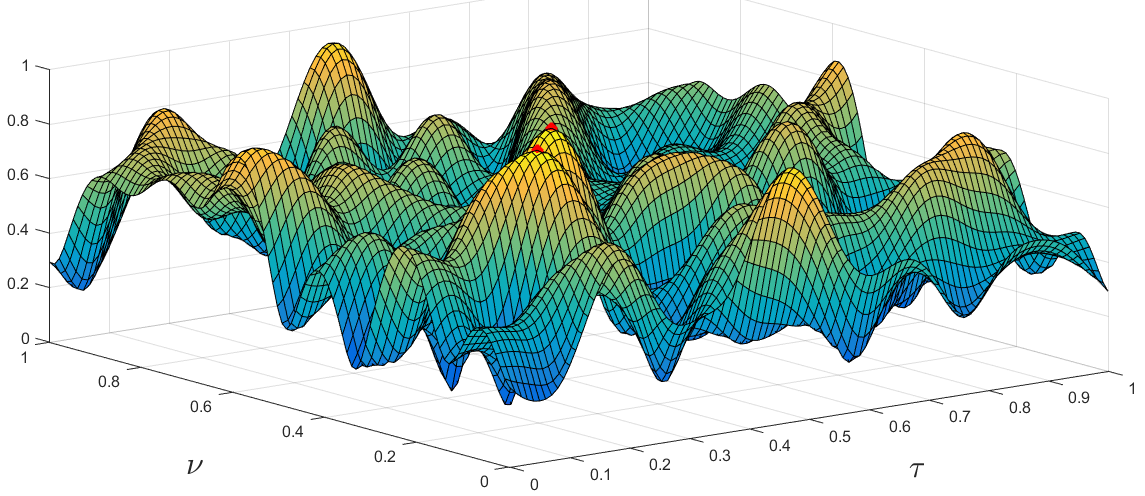}
	\end{subfigure}
	\caption{{\color{\change}Noiseless case. The true sources are located at $\bm{r}_1=(0.2,0.2)$ and $\bm{r}_2=(0.3,0.3)$. We set $N=4$ and solve \eqref{eq.sdp}. Left and Right images show $\|\bm{q}(\bm{r})\|_2$ for SMV ($R=1$) and MMV ($R=30$) cases, respectively. Red markers show where $\|\bm{q}(\bm{r})\|_2$ equals one.}}  
	\label{fig2}
\end{figure}

\begin{figure}[t]
	
	\begin{subfigure}[b]{.5\textwidth}
		\centering
		\includegraphics[scale=.4]{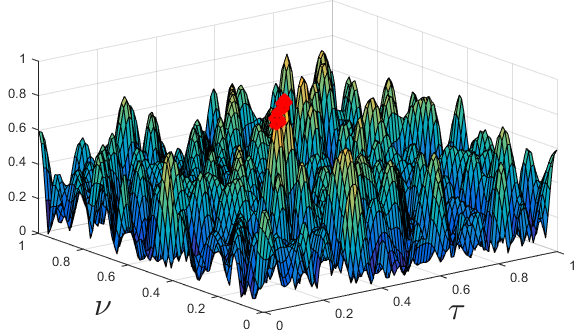}
	\end{subfigure}
	\begin{subfigure}[b]{.5\textwidth}
		\centering
		\includegraphics[scale=.4]{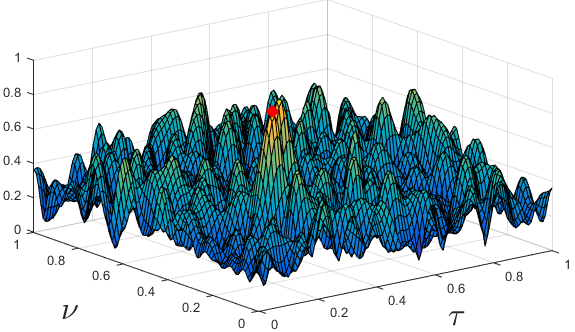}
	\end{subfigure}
	\caption{ {\color{\change}Noiseless case. The true sources are located at $\bm{r}_1=(0.2,0.2)$ and $\bm{r}_2=(0.3,0.3)$. We set $N=8$ and solve \eqref{eq.sdp}. Left and Right images show $\|\bm{q}(\bm{r})\|_2$ for SMV ($R=1$) and MMV ($R=20$) cases, respectively. Red markers show where $\|\bm{q}(\bm{r})\|_2$ achieves one}.}
	
	\label{fig4}
\end{figure}

\begin{figure}[t]
	
	\begin{subfigure}[b]{.5\textwidth}
		\centering
		\includegraphics[scale=.2]{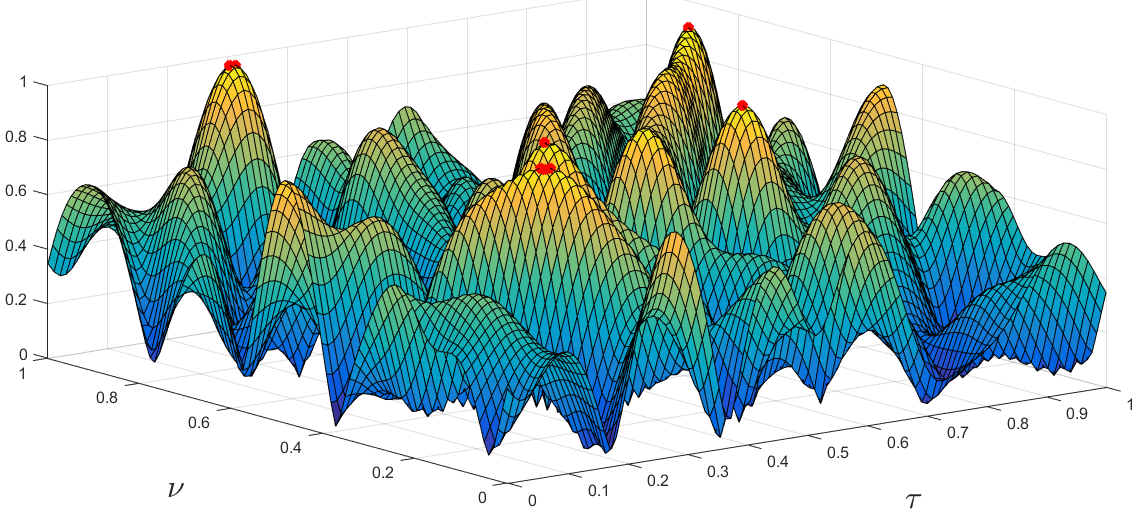}
	\end{subfigure}
	\begin{subfigure}[b]{.5\textwidth}
		\centering
		\includegraphics[scale=.2]{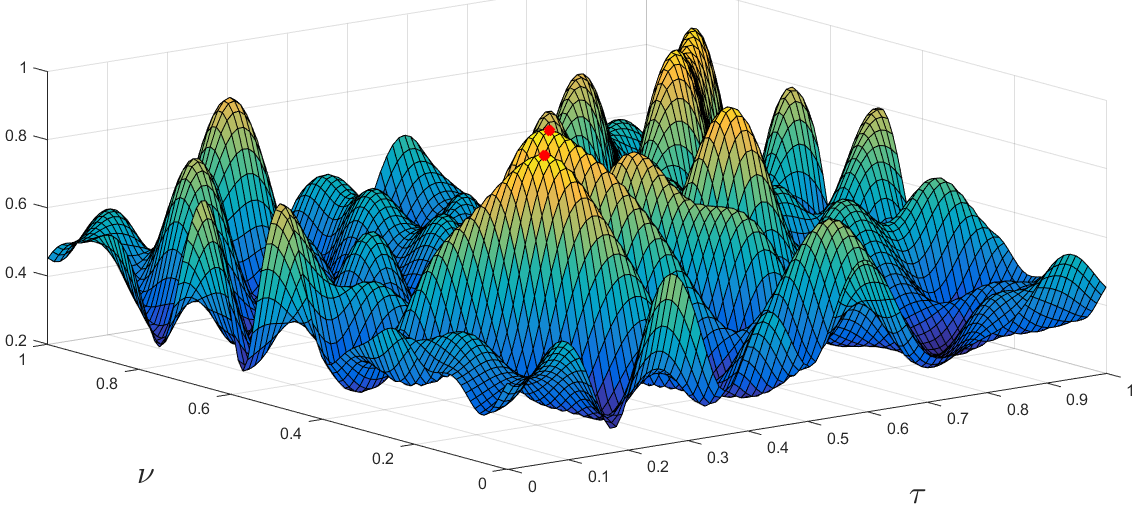}
	\end{subfigure}
	\caption{{\color{\change}Noisy case. The true sources are located at $\bm{r}_1=(0.2,0.2)$, $\bm{r}_2=(0.3,0.3)$. We set $N=4, \eta=0.8, {\rm SNR}=10{\rm dB}$ and solve \eqref{sdp_noisy}. Left and Right images show $\|\bm{q}(\bm{r})\|_2$ for SMV ($R=1$) and MMV ($R=50$) cases, respectively. Red markers show where $\|\bm{q}(\bm{r})\|_2$ equals one.}}
	\label{fig3}
\end{figure}

\begin{figure}[t]
		\centering
		\includegraphics[scale=.5]{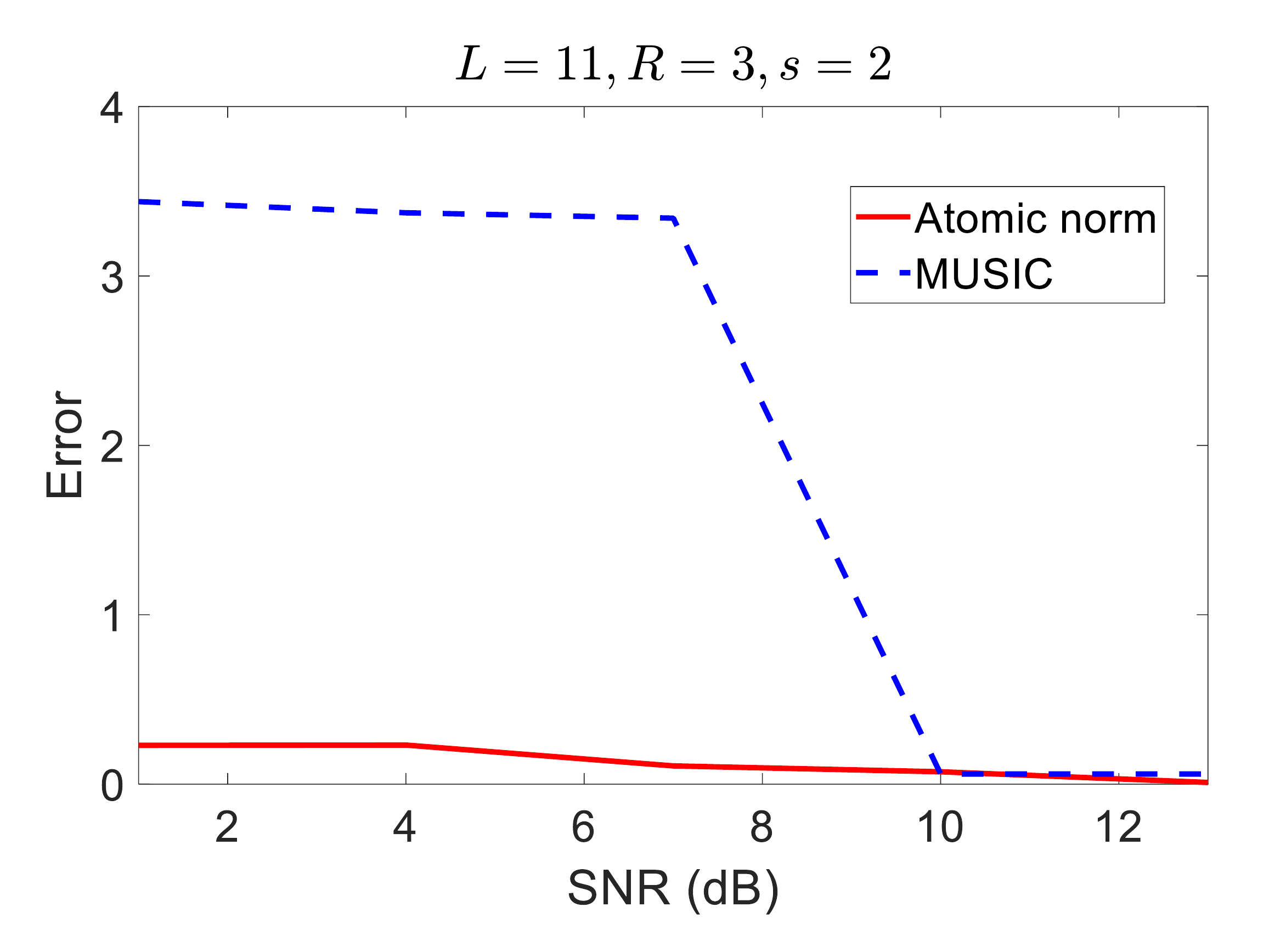}
	\caption{{\color{\change}This figure compares our atomic norm approach with MUSIC and depicts the error \eqref{eq.error} as a function of SNR with parameters $R=3, L=11, s=2$.} }	
	\label{fig.music}
\end{figure}

In this section, we consider the benefits of using MMVs in both noiseless and noisy cases. Let the probing signal $x_l, ~l=-N,...,N$ and the coefficients $\bm{b}_{jm}$ be drawn from i.i.d. uniform distribution on the complex unit sphere with parameter $N=4$. First, in the left and right images of Figure \ref{fig1}, we consider two time-frequency shifts in locations {\color{\change} $\bm{r}_1=(0.2,0.8)$ and $\bm{r}_2=(0.5,0.5)$} with $R=1$ and $R=10$, respectively. We implement the problem \eqref{eq.sdp} via SDPT3 in CVX package \cite{cvx} and plot $\|\bm{q}(\bm{r})\|_2$ where $\bm{q}(\bm{r})$ is obtained from \eqref{eq.dual_pol}. Then, according to \eqref{eq.support_estimates}, we estimate time-frequency shifts by checking locations where $\|\bm{q}(\bm{r})\|_2$ achieves one. As it turns out from the left and right images of Figure \ref{fig1}, while the SMV case ($R=1$) fails and finds spurious sources, the MMV case ($R=10$) localize the delay-Doppler pairs correctly. This in turn shows that using more antenna arrays improves the recovery performance for a fixed and weak time-frequency minimum separation. In the second experiment, we check a case where the sources are closer to each other as shown in Figure \ref{fig2}. Again, we can see the superiority of using MMV ($R=30$) over SMV. All in all, one can infer from Figures \ref{fig1} and \ref{fig2} that under a fixed number of measurements $N=4$, benefiting more MMVs can make the required minimum separation condition weaker (alternatively leading to a more relaxed condition).  We also investigate a similar noiseless problem for $N=8$ and evaluate the performance of our MMV method over the SMV case in Figure \ref{fig4} and again observe the superiority of MMV over SMV. In another experiment, we examine the noisy case where we consider a complex noise $\bm{W}$ with i.i.d. Gaussian elements such that the signal to noise ratio (SNR) defined as {\color{\change}$\text{SNR}=10\log(\tfrac{\|\bm{Y}\|_{F}^{2}}{\|\bm{W}\|_{F}^{2}})$ } is equal to $10\,$dB.
We choose two delay-Doppler pairs $\bm{r}_1=(0.2,0.8)$ and $\bm{r}_2=(0.5,0.5)$ and implement \eqref{sdp_noisy} {\color{\change}with $\eta=\frac{8}{3}\|\bm{W}\|_F$}. As shown in the left and right images of Figure \ref{fig3}, increasing the array elements (from $R=1$ in the left image to $R=50$ in the right image) can improve the recovery performance of delay-Doppler pairs. {\color{\change}In the last experiment, we compare our algorithm with the MUSIC approach \cite{music}. We investigate the performance against different noise levels by using the error defined by
	\begin{align}\label{eq.error}
{\rm Error}:=\frac{1}{s}\sum_{i=1}^{s}L\sqrt{(\widehat{\tau}_i-\tau)^2+(\widehat{\nu}_i-\nu_i)^2},
	\end{align}
where $(\widehat{\tau}_i,\widehat{\nu}_i)$ are the estimated time-frequency shifts. As is shown in Figure \ref{fig.music}, while MUSIC yields to large errors in low SNR regimes, our atomic norm approach successfully recovers the true time-frequency shifts. } 


{\color{\change}\section{Conclusion}
In this paper, we addressed the estimation of continuous time-frequency shifts of a known waveform from a small number of samples in the presence of MMVs. We converted this problem as a 2-D atomic norm minimization which then was relaxed as a tractable SDP. We showed that the unique recovery of continuous parameters is possible as long as the time-freq. shifts satisfy a minimum separation condition and that the number of samples per measurement vector is linearly related to the number of time-freq. shifts. The effectiveness of our approach over the SMV case and MUSIC method was certified using various numerical experiments.
}
\section{Appendix}
\subsection{Equivalence of \eqref{eq.rel1} and \eqref{eq.sampled_y} }\label{proof.equivalence}
{\color{black}We provide a proof for this equivalence. Similar proofs of such analog to discrete conversion are also provided in \cite[Appendix A]{heckel2016super} and \cite[Appendix A]{suliman2018blind}.
}
By sampling at rate $\tfrac{1}{B}$, the equation \eqref{eq.rel1} can be displayed as follows:
\begin{align}\label{24}
&y_m(\tfrac{p}{B})=\sum_{j=1}^sb_{jm}x(\tfrac{p}{B}-\overline{\tau}_j){\rm e}^{i2\pi \overline{\nu}_j \tfrac{p}{B}}, p=-N,..., N,
\end{align}
we know that (by applying the discrete Fourier transform (DFT) and inverse DFT (IDFT) to $x$):
\begin{align}\label{25}
&x(\tfrac{p}{B}-\tfrac{\tau_jTB}{B})=x(\tfrac{p}{B}-\tfrac{\tau_jL}{B})=x[p-\tau_jL]=\tfrac{1}{L}\sum_{k=-N}^{N}\Big(\sum_{l=-N}^{N}x[l]e^{-\tfrac{i2\pi kl}{L}}\Big){\rm e}^{-\tfrac{i2\pi \tau_jLk}{L}}e^{\tfrac{i2\pi kp}{L}},
\end{align}
{\color{\change}where $x[n]:=x(\frac{n}{B})$ for some integer $n$.}
Substituting \eqref{25} into \eqref{24} leads to
\begin{align}\label{eq.rel6}
&y(\tfrac{p}{B})=\tfrac{1}{L}\sum_{j=1}^{s}b_{jm}\sum_{k=-N}^{N}\Big(\sum_{l=-N}^{N}x[l]e^{-\tfrac{i2\pi kl}{L}}\Big){\rm e}^{-\tfrac{i2\pi \tau_jLk}{L}}\nonumber\\&{\rm e}^{\tfrac{i2\pi kp}{L}}{\rm e}^{i2\pi {\nu}_jp}=\sum_{j=1}^{s}b_{jm}\tfrac{1}{L}\sum_{k,l=-N}^{N}x[l]{\rm e}^{\tfrac{i2\pi k(p-l)}{L}}\nonumber\\
&{\rm e}^{i2\pi [p\nu_j-k\tau_j]}=\sum_{j=1}^{s}b_{jm}{\rm e}^{i2\pi p \nu_j }\tfrac{1}{L}\sum_{l=-N}^{N}\sum_{k=-N}^{N}x[l]\nonumber\\
&{\rm e}^{i2\pi [\tfrac{p-l}{L}-\tau_j]k}=\sum_{j=1}^{s}b_{jm}{\rm e}^{i2\pi p\nu_j}\tfrac{1}{L}\sum_{n=p-N}^{p+N}\sum_{k=-N}^{N}x[p-n]\cdot\nonumber\\
&{\rm e}^{i2\pi [\tfrac{n}{L}-\tau_j]k} ~~~ p=-N,..., N, ~ m=1,..., R.
\end{align}
By using the definition \eqref{eq.diric}, the fact that
\begin{align}
\sum_{k=-N}^{N}D_N(\tfrac{k}{L}-\nu_j){\rm e}^{\tfrac{i2\pi pk}{L}}={\rm e}^{i2\pi p\nu_j},
\end{align}
and the periodicity property of $x_l$, \eqref{eq.rel6} becomes
	\begin{align}
	&y_m(\tfrac{p}{B})=\sum_{j=1}^{s}b_{jm}{\rm e}^{i2\pi p\nu_j}\tfrac{1}{L}\sum_{n=-N}^{N}\sum_{k=-N}^{N}x[n-p]{\rm e}^{i2\pi [\tfrac{n}{L}-\tau_j]k}\nonumber\\
	&=\sum_{j=1}^{s}b_{jm}{\rm e}^{i2\pi p\nu_j}\sum_{n=-N}^{N}x[n-p]D_N(\tfrac{n}{L}-\tau_j)\nonumber\\
	&=\sum_{j=1}^{s}b_{jm}\sum_{k=-N}^{N}D_N(\tfrac{k}{L}-\nu_j){\rm e}^{\tfrac{i2\pi pk}{L}}\sum_{n=-N}^{N}x[n-p]\nonumber\\
	&\cdot D_N(\tfrac{n}{L}-\tau_j)=\nonumber\\
	&\sum_{j=1}^{s}b_{jm}\sum_{k,n=-N}^{N}D_N(\tfrac{k}{L}-\nu_j)D_N(\tfrac{n}{L}-\tau_j)x_{n-p}{\rm e}^{\tfrac{i2\pi pk}{L}}.
	\end{align}
\section{Proof of Proposition \ref{prop.optimality}}\label{proof.propositon_optimal}
We begin with $\bm{\Lambda}$ which lies in the feasible set of \eqref{eq.dual_prob}, since due to \eqref{eq.dualnorm_def} and the assumptions \eqref{dual}, we have:
\begin{align}\label{rel5}
&\|\bm{G}^{\rm H}\bm{\Lambda}\|_{\mathcal{A}}^d=\sup_{\substack{   \bm{r}\in[0,1]^2}}\|\bm{q}(\bm{r})\|_2 \le 1.
\end{align}
We proceed by writing
\begin{align}
&\|\bm{X}\|_{\mathcal{A}}\ge \|\bm{G}^{\rm H}\bm{\Lambda}\|_{\mathcal{A}}^d\|\bm{X}\|_{\mathcal{A}}\stackrel{(\RN{1})}{\ge}{\rm Re}~\langle \bm{G}^{\rm H}\bm{\Lambda}, \bm{X} \rangle_F={\rm Re}~\langle \bm{\Lambda}, \bm{Y} \rangle_F=\nonumber\\
&{\rm Re}~\langle \bm{\Lambda}, \bm{G} \sum_{\bm{r}_j\in\mathcal{S}}c_j\bm{a}(\bm{r}_j)\bm{\varphi}_j^{\rm H} \rangle_F=\sum_{\bm{r}_j\in\mathcal{S}}{\rm Re}~c_j\langle \bm{\varphi}_j, \bm{\varphi}_j \rangle
\stackrel{(\RN{2})}{=}\nonumber\\
&\sum_{\bm{r}_j\in\mathcal{S}}c_j=\|\bm{X}\|_{\mathcal{A}},
\end{align}
where the inequality $(\RN{1})$ is due to \eqref{rel5}, and the equality $(\RN{2})$ stems from the assumptions in \eqref{dual}. The latter relation shows that all the inequalities must be turned into equality. Thus, ${\rm Re}\langle \bm{\Lambda},\bm{Y}\rangle=\|\bm{X}\|_{\mathcal{A}}$ which in turn shows that $(\bm{X},\bm{\Lambda})$ are primal-dual optimal solutions. For uniqueness, we argue by contradiction and assume that there exists another optimal primal solution $\widehat{\bm{X}}=\sum_{\bm{r}_j\in\widehat{\mathcal{S}}}\widehat{c}_j\bm{a}(\bm{r}_j)\widehat{\bm{\varphi}}_j^{\rm H}$ where $\widehat{S}\neq \mathcal{S}$. It holds that
\begin{align}
&\|\widehat{\bm{X}}\|_{\mathcal{A}}={\rm Re}~\langle \bm{\Lambda}, \bm{G}\widehat{\bm{X}}\rangle={\rm Re}~\langle \bm{\Lambda}, \bm{G} \sum_{\bm{r}_j\in\widehat{\mathcal{S}}}\widehat{c}_j\bm{a}(\bm{r}_j)\widehat{\bm{\varphi}}_j^{\rm H}\rangle=\nonumber\\
&\sum_{\bm{r}_j\in {\mathcal{S}}}{\rm Re}~\widehat{c}_j\langle \widehat{\bm{\varphi}}_j, \bm{q}(\bm{r}_j)\rangle+
\sum_{\bm{r}_j\in \widehat{\mathcal{S}}\setminus \mathcal{S}}{\rm Re}~\widehat{c}_j\langle \widehat{\bm{\varphi}}_j, \bm{q}(\bm{r}_j)\rangle
<\sum_{\bm{r}_j\in\widehat{\mathcal{S}}}\widehat{c}_j\nonumber\\
&=\|\widehat{\bm{X}}\|_{\mathcal{A}},
\end{align}
where we used the assumptions in \eqref{dual} in the last inequality. Hence, we have a contradiction and $\widehat{\mathcal{S}}=\mathcal{S}$. As a consequence, since $\bm{a}(\bm{r}_j),~ \bm{r}_j\in\mathcal{S}$ are linearly independent, the optimal primal solution is unique. 
\subsection{Proof of Theorem \ref{thm.main}}\label{construction.dual.poly}
In this section, we prove Theorem \ref{thm.main} by constructing 2D vector-valued dual polynomial $\bm{q}$ satisfying \eqref{dual}. Without loss of generality, we assume that $N$ is even and define the squared Fejer kernel
\begin{align}
K(t):=\tfrac{1}{M}\sum\limits_{k=-N}^{N}{{{g}_{k}}{{\rm e}^{i2\pi tk}}}:={{\left( \tfrac{\sin (M\pi t)}{M\sin (\pi t)} \right)}^{4}},\text{  }M=\tfrac{N}{2}+1,
\end{align}
where $g_k$ is the discrete convolution of two triangular functions.
First, in the following, we construct a deterministic dual polynomial satisfying \eqref{dual} which is later used in our analysis:

\begin{align}\label{determin}
& \overline{\bm{q}}(\mathbf{r)=}\sum\limits_{j=1}^{s}{{ \overline{\bm{\alpha} }}_{j}}{ \overline{G}(\bm{r}-{{\bm{r}}_{j}})}+{{ \overline{\bm{\beta }}}_{j}}{{ \overline{G}}^{(1,0)}}(\bm{r}-{{\bm{r}}_{j}}) +\nonumber \\
& \overline{\bm{\gamma }}_{j}{{ \overline{G}}^{(0,1)}}(\bm{r}-{{\bm{r}}_{j}}),
\end{align}
where the coefficients ${{ \overline{\bm{\alpha} }}_{j}},{{ \overline{\bm{\beta} }}_{j}},{{ \overline{\bm{\gamma} }}_{j}}\in {{\mathbb{C}}^{R\times 1}}$, $\overline{G}^{(m,n)} = \tfrac{\partial^{m}\partial^{n}\overline{G}}{\partial \tau^{m}\partial \nu^{n}} $ and $ \overline{G}(\bm{r}):=K(\tau )K(\nu )$. 
An important requirement for the condition \eqref{dual} to hold, is that $ \overline{\bm{q}}(\bm{r})$ reaches the local maxima by choosing the specific coefficients ${{ \overline{\bm{\alpha}}}_{j}}, {{ \overline{\bm{\beta} }}_{j}}, {{ \overline{\bm{\gamma}}}_{j}}$ satisfying
\begin{align}\label{eq.conditions_qbar}
& \overline{\bm{q}}({\bm{r}_j})=\bm{\varphi}_j, & \forall \bm{r}_j \in \mathcal{S} \nonumber\\ & \overline{\bm{q}}^{(1,0)}({\bm{r}_j})=\bm{0}\in \mathbb{C}^{R\times 1},  &\forall \bm{r}_j \in \mathcal{S}\nonumber\\
& \overline{\bm{q}}^{(0,1)}({\bm{r}_j})=\bm{0}\in \mathbb{C}^{R\times 1},  &\forall \bm{r}_j \in \mathcal{S},
\end{align} 
where ${{ \overline{\bm{q}}}^{(m,n)}}(\bm{r}):=\tfrac{\partial^{m}\partial^{n}\overline{\bm{q}}}{\partial \tau^{m}\partial \nu^{n}} $. 
Now, we construct the random polynomial $\bm{q}(\bm{r})$ with function $G_{(m,n)}(\bm{r},\bm{r}_j),m,n=0,1$ as
\begin{align}\label{dual-poly}
&\bm{q}(\mathbf{r)=}\sum\limits_{j=1}^{s}{{\bm{\alpha} }_{j}}{{{G}_{^{(0,0)}}}(\bm{r},{{\bm{r}}_{j}})}+{{\bm{\beta} }_{j}}{{G}_{^{(1,0)}}}(\bm{r},{{\bm{r}}_{j}})\nonumber\\
&+ {{\bm{\gamma} }_{j}}{{G}_{^{(0,1)}}}(\bm{r},{{\bm{r}}_{j}}),
\end{align}
where the coefficients $\bm{\alpha}_k,\bm{\beta}_k,\bm{\gamma}_k$ are such that:
\begin{align}\label{shart}
&{\bm{q}}({\bm{r}_j})=\bm{\varphi}_j,\quad\quad\quad\quad\quad~~ \forall \bm{r}_j \in \mathcal{S} \nonumber\\ &{\bm{q}}^{(1,0)}({\bm{r}_j})=\bm{0}\in \mathbb{C}^{R\times 1},\quad \forall \bm{r}_j \in \mathcal{S}\nonumber\\
&{\bm{q}}^{(0,1)}({\bm{r}_j})=\bm{0}\in \mathbb{C}^{R\times 1},\quad \forall \bm{r}_j \in \mathcal{S}
\end{align}
and $\overline{G}:=\mathds{E}G$ is the expectation of kernel $G$.
The dual polynomial $\bm{q}$ in \eqref{dual-poly} can be regarded as a random version of $\overline{\bm{q}}$ in \eqref{determin} with the randomness introduced by $\bm{x}$.
\subsubsection{Choice of coefficients}
We choose the coefficients ${{\bm{\alpha} }_{j}},{{\bm{\beta} }_{j}},{{\bm{\gamma} }_{j}}$ to construct $\bm{q}({\bm r})$ such that \eqref{shart} holds with high probability. 
Writing \eqref{determin} in matrix form, yields
\begin{align}
\underbrace{\left[ \begin{matrix}
	{{ \overline{\bm{D}}}^{(0,0)}} & \kappa^{-1}{{ \overline{\bm{D}}}^{(1,0)}} & \kappa^{-1}{{ \overline{\bm{D}}}^{(0,1)}}  \\
	-\kappa^{-1}{{ \overline{\bm{D}}}^{(1,0)}} & -\kappa^{-2}{{ \overline{\bm{D}}}^{(2,0)}} & -\kappa^{-2}{{ \overline{\bm{D}}}^{(1,1)}}  \\
	-\kappa^{-1}{{ \overline{\bm{D}}}^{(0,1)}} & -\kappa^{-2}{{ \overline{\bm{D}}}^{(1,1)}} & -\kappa^{-2}{{ \overline{\bm{D}}}^{(0,2)}}  \\
	\end{matrix} \right]}_{ \overline{\bm{D}}}\left[ \begin{matrix}
\overline{\bm{\alpha} }  \\
\kappa \overline{\bm{\beta} }  \\
\kappa \overline{\bm{\gamma} }  \\
\end{matrix} \right]=\left[ \begin{matrix}
\overline{\bm{\Phi}}   \\
\bm{0}  \\
\bm{0}  \\
\end{matrix} \right],
\end{align}
where $(\kappa^{2}=\left| {K}''(0) \right|=\sqrt{\tfrac{\pi^2}{3}(N^2+4N)},K(0)=1)$ and ${{\left[ {{ \overline{\bm{D}}}^{(m,n)}} \right]}_{\kappa,j}}:={{ \overline{G}}^{(m,n)}}({{\bm{r}}_{\kappa}}-{{\bm{r}}_{_{j}}})$, $\overline{\bm{\Phi}}=[\bm{\overline{\varphi}}_1, \bm{\overline{\varphi}}_2,\hdots,\bm{\overline{\varphi}}_s]^{\top} \in \mathbb{C}^{s\times R}$ , $\bm{\overline{\alpha}}=[\bm{\overline{\alpha}}_{1},\bm{\overline{\alpha}}_{2},\hdots,\bm{\overline{\alpha}}_{s}]^{\top} \in \mathbb{C}^{s\times R}$, $\bm{\overline{\beta}}=[\bm{\overline{\beta}}_{1}, \bm{\overline{\beta}}_{2},\hdots, \bm{\overline{\beta} }_{s}]^{\top} \in \mathbb{C}^{s\times R}$,
$\bm{\overline{\gamma}}=[\bm{\overline{\gamma}}_1,\bm{\overline{\gamma}}_2,\hdots,\bm{\overline{\gamma}}_s]^{\top}\in \mathbb{C}^{s\times R}$ and
$ \overline{\bm{D}}$ is symmetric because $ \overline{\bm{D}}^{(0,0)},~\overline{\bm{D}}^{(1,1)}, \overline{\bm{D}}^{(2,0)},~\overline{\bm{D}}^{(0,2)}$ are symmetric and $ \overline{\bm{D}}^{(1,0)},~\overline{\bm{D}}^{(0,1)}$ are antisymmetric. ${ \overline{\bm{D}}}$ is invertible and also the coefficients can be obtained as
\begin{align}
\left[ \begin{matrix}
\overline{\bm{\alpha} }  \\
\kappa \overline{\bm{\beta} }  \\
\kappa \overline{\bm{\gamma} }  \\
\end{matrix} \right]={{ \overline{\bm{D}}}^{-1}}\left[ \begin{matrix}
\bm{\Phi}   \\
\bm{0 } \\
\bm{0 }  \\
\end{matrix} \right]= \overline{\mathbf{L}}\bm{\Phi},
\end{align} 
where $ \overline{\mathbf{L}}\in \mathbb{C}^{3s\times s}$  is the first $ s $ columns of ${ \overline{\bm{D}}^{-1}}$.
\begin{prop}\cite[Proposition 8.2]{heckel2016super}
	${{ \overline{\bm{D}}}}$ is invertible and
	\begin{align}
	\left\| \mathbf{I}- \overline{\bm{D}} \right\|\le 0.19808,
	\end{align}
	\begin{align}
	\left\|  \overline{\bm{D}} \right\|\le 1.19808,
	\end{align}
	\begin{align}
	\left\| {{ \overline{\bm{D}}}^{-1}} \right\|\le 1.24700.
	\end{align}	
\end{prop} 
Next, we choose the coefficients ${{\bm{\alpha} }},{{\bm{\beta} }},{{\bm{\gamma} }}\in {{\mathbb{C}}^{s\times R}}$ such that the conditions \eqref{shart} hold. First, write \eqref{dual-poly} in matrix form as
\begin{align}\label{36}
\underbrace{\left[ \begin{matrix}
	\bm{D}_{(0,0)}^{(0,0)} & \kappa^{-1}\bm{D}_{(1,0)}^{(0,0)} & \kappa^{-1}\bm{D}_{(0,1)}^{(0,0)}  \\
	-\kappa^{-1}\bm{D}_{(0,0)}^{(1,0)} & -\kappa^{-2}\bm{D}_{(1,0)}^{(1,0)} & -\kappa^{-2}\bm{D}_{(0,1)}^{(1,0)}  \\
	-\kappa^{-1}\bm{D}_{(0,0)}^{(0,1)} & -\kappa^{-2}\bm{D}_{(1,0)}^{(0,1)} & -\kappa^{-2}\bm{D}_{(0,1)}^{(0,1)}  \\
	\end{matrix} \right]}_{\bm{D}}\left[ \begin{matrix}
\bm{\alpha}   \\
{{\kappa}}\bm{\beta}   \\
{{\kappa}}\bm{\gamma}   \\
\end{matrix} \right]=\left[ \begin{matrix}
\bm{\Phi}   \\
\bm{0 }  \\
\bm{0 }  \\
\end{matrix} \right],
\end{align}
where  $\Big[\bm{D}^{(m,n)}_{(m',n')}\Big]_{j,k}
:=G^{(m,n)}_{(m',n')}(\bm{r}_j,\bm{r}_k)$, $\bm{\Phi}=[\bm{\varphi}_1, \bm{\varphi}_2,\hdots,\bm{\varphi}_s]^{\top} \in \mathbb{C}^{s\times R}$ , $\bm{\alpha}=[\bm{\alpha}_{1},\bm{\alpha}_{2},\hdots,\bm{\alpha}_{s}]^{\top}$, $\bm{\beta}=[\bm{\beta}_{1}, \bm{\beta}_{2},\hdots, \bm{\beta}_{s}]^{\top}$, $\bm{\gamma}=[\bm{\gamma}_1,\bm{\gamma}_2,\hdots,\bm{\gamma}_s]^{\top}$, where $\bm{\alpha}_j$, $\bm{\beta}_j$, $\bm{\gamma}_j \in \mathbb{C}^{R \times 1}$ with $j={1,\hdots, s}$.
To prove the existence of coefficients $\bm{\alpha},\bm{\beta}, \bm{\gamma}$, we show that $\bm{D}$ in \eqref{36} is invertible with high probability. Define the event
\begin{align}
{{\zeta }_{\xi }}=\{\left\| \bm{D}-\overline{\bm{D}} \right\|\le \xi \}.
\end{align}
If ${{\zeta }_{\xi }}$ occurs with $\xi \in \left( 0,\tfrac{1}{4} \right]$, $\bm{D}$ is invertible since
\begin{align}
\left\| \mathbf{I}-\bm{D} \right\|\le \left\| \bm{D}- \overline{\bm{D}} \right\|+\left\|  \overline{\bm{D}}-\mathbf{I} \right\|\le \xi +0.1908\le 0.4408.
\end{align}
Hence, $\bm{\alpha}_j,\bm{\beta}_j,\bm{\gamma}_j$ can be given as
\begin{align}\label{coefficient}
\left[ \begin{matrix}
{\bm{\alpha} }  \\
\kappa{\bm{\beta} }  \\
\kappa{\bm{\gamma} }  \\
\end{matrix} \right]={{{\bm{D}}}^{-1}}\left[ \begin{matrix}
\bm{\Phi}   \\
\bm{0 }  \\
\bm{0 }  \\
\end{matrix} \right]={\mathbf{L}}\bm{\Phi},
\end{align}
where $\mathbf{L} \in \mathbb{C}^{3s\times s}$ is the first $s$ columns of $\bm{D}^{-1}$.
To proceed, we use the following important lemma about the concentration of $\bm{L}$ around $\overline{\bm{L}}$:
\begin{lem}\label{lem.L_bounds}(\cite[Lemma 8.4]{heckel2016super})
	If the event ${{\zeta }_{\xi }}$ with $\xi  \in (0,\tfrac{1}{4}]$ occurs, then we have
	\begin{align}
	\| \mathbf{L} \|\le 2.5, ~\| \mathbf{L}- \overline{\mathbf{L}}\|\le 2.5\xi. 
	\end{align}
\end{lem}
The following lemma provides conditions that ${\zeta }_{\xi }$ occurs with high probability. We use this lemma to complete our proof.
\begin{lem}\label{2}(\cite[Lemma 8.6]{heckel2016super})
	If
	\begin{align}
	L\ge s\tfrac{{{c}_{1}}}{{{\xi }^{2}}}{{\log }^{2}}\tfrac{18{{s}^{2}}}{\delta},
	\end{align}
	then,
	\begin{align}
	\mathds{P} [{{\zeta }_{\xi }}] \ge 1-\delta .\end{align}  
	
\end{lem}

\subsubsection{Showing that \texorpdfstring{$\bm{q}(\bm{r})  $}{TEXT} and \texorpdfstring{$\overline{\bm{q}}(\bm{r}) $}{TEXT} are close on a grid }
The goal of this section is to show that $\bm{q}(\bm{r})\text{ }$ and $\overline{\bm{q}}(\bm{r})\text{ }$ are close in a set of grid points $\Omega$.
{\color{\change}
\begin{lem}\label{3}
	Let $\Omega \subset {{[0,1]}^{2}}$ be a finite set of points. Fix $0<\epsilon \le 1$ and $\delta {>0}$. If
		\begin{align}
		L \ge cs \max\{\log^2{\tfrac{12sL}{\delta}}
		(1+\tfrac{1}{R}\log(\tfrac{|\Omega|}{\delta})),\log^2\tfrac{18s^2}{\delta}(1+\tfrac{1}{R}\log(\tfrac{|\Omega|}{\delta}))\}
		\end{align}
	then,
	$\mathds{P}\left[ \underset{\bm{r}\in \Omega }{\mathop{\max}}\,\tfrac{1}{\kappa^{m+n}}{{\left\| {{\bm{q}}^{(m,n)}}(\bm{r})-{{\overline{\bm{q}}}^{(m,n)}}(\bm{r}) \right\|}_{2}}\le \epsilon  \right]\ge 1-4\delta $.
\end{lem}}
It is straightforward to verify that $(m,n)$-th partial derivative of the dual polynomial $\bm{q}(\bm{r})$ can be written as
\begin{align}\label{45}
&\tfrac{1}{\kappa^{m+n}}{{\bm{q}}^{(m,n)}}(\bm{r})=\sum\limits_{j=1}^{s}{G_{(0,0)}^{(m,n)}}(\bm{r},{{\bm{r}}_{j}}){{\bm{\alpha} }_{j}}+\tfrac{1}{{{\kappa}}}G_{(1,0)}^{(m,n)}(\bm{r},{{\bm{r}}_{j}}){{\kappa}}{{\bm{\beta} }_{j}}+\tfrac{1}{{{\kappa}}}G_{(0,1)}^{(m,n)}(\bm{r},{{\bm{r}}_{j}}){{\kappa}}{{\bm{\gamma} }_{j}}\nonumber \\ &={{({{\bm{w}}^{(m,n)}}(\bm{r}))}^{\rm H}}\mathbf{L}\bm{\Phi},
\end{align}
where
\begin{align}
&{{\text{(}{{\bm{w}}^{(m,n)}})}^{\rm H}}(\bm{r}):=\tfrac{1}{\kappa^{m+n}}[G_{(0,0)}^{(m,n)}(\bm{r},{{\bm{r}}_{1}}),...,G_{(0,0)}^{(m,n)}(\bm{r},{{\bm{r}}_{s}}),\nonumber\\
&\tfrac{1}{{\kappa}}G_{(1,0)}^{(m,n)}(\bm{r},{{\bm{r}}_{1}}),...,\tfrac{1}{{{\kappa}}}G_{(1,0)}^{(m,n)}(\bm{r},{{\bm{r}}_{s}}),\nonumber\\
&\tfrac{1}{{{\kappa}}}G_{(0,1)}^{(m,n)}(\bm{r},{{\bm{r}}_{1}}),...,\tfrac{1}{{{\kappa}}}G_{(0,1)}^{(m,n)}(\bm{r},{{\bm{r}}_{s}})].
\end{align}    
Due to $\mathds{E}\left[ G_{({m}',{n}')}^{(m,n)}(\bm{r},{{\bm{r}}_{j}}) \right]={{ \overline{G}}^{(m+{m}',n+{n}')}}(\bm{r}-{{\bm{r}}_{j}})$, it holds that $\mathds{E}\left[ {{\bm{w}}^{(m,n)}}(\bm{r}) \right]={{ \overline{\bm{w}}}^{(m,n)}}(\bm{r})$, where
\begin{align}
&{{\text{(}{{ \overline{\bm{w}}}^{(m,n)}})}^{\rm H}}(\bm{r}):=\tfrac{1}{\kappa^{m+n}}[{{ \overline{G}}^{(m,n)}}(\bm{r}-{{\bm{r}}_{1}}),...,{{ \overline{G}}^{(m,n)}}(\bm{r}-{{\bm{r}}_{s}}),\nonumber\\
&\tfrac{1}{{{\kappa}}}{{ \overline{G}}^{(m+1,n)}}(\bm{r}-{{\bm{r}}_{1}}),...,\tfrac{1}{{{\kappa}}}{{ \overline{G}}^{(m+1,n)}}(\bm{r}-{{\bm{r}}_{s}}),\nonumber\\
&\tfrac{1}{{{\kappa}}}{{ \overline{G}}^{(m,n+1)}}(\bm{r}-{{\bm{r}}_{1}}),...,\tfrac{1}{{{\kappa}}}{{ \overline{G}}^{(m,n+1)}}(\bm{r}-{{\bm{r}}_{s}})].
\end{align}
Now, we can decompose \eqref{45} as follows
\begin{align}
&\tfrac{1}{\kappa^{m+n}}{{\bm{q}}^{(m,n)}}(\bm{r})=(\bm{w}^{(m,n)})^{{\rm H}}(\bm{r})\mathbf{L}\bm{\Phi}={{\text{(}{{ \overline{\bm{w}}}^{(m,n)}})}^{\rm H}}(\bm{r}) \overline{\mathbf{L}}\bm{\Phi}\nonumber\\ &-\underbrace{{{({{\bm{w}}^{(m,n)}}(\bm{r})-{{ \overline{\bm{w}}}^{(m,n)}})}^{\rm H}}\mathbf{L}\bm{\Phi} }_{\bm{I}_{1}^{(m,n)}(\bm{r})}+\underbrace{{{\text{(}{{ \overline{\bm{w}}}^{(m,n)}}\text{)}}^{\rm H}}(\mathbf{L}- \overline{\mathbf{L}})\bm{\Phi} }_{\bm{I}_{2}^{(m,n)}(\bm{r})}\nonumber\\
&=\tfrac{1}{\kappa^{m+n}}{{ \overline{\bm{q}}}^{(m,n)}}(\bm{r})+{\bm{I}_{1}^{(m,n)}(\bm{r})}+{\bm{I}_{2}^{(m,n)}(\bm{r})}.
\end{align}
The following lemmas show that ${\bm{I}_{1}^{(m,n)}(\bm{r})}$ and ${\bm{I}_{2}^{(m,n)}(\bm{r})}$ are small on a set of grid points $\Omega $ with high probability. {\color{\chang} To show this, we benefit from a generalized Bernstein inequality which is proposed in \cite[Lemma 4]{yang2018sample}. In what follows, we observe that using this inequality makes the lower bound on the sample complexity $L$ decrease as $R$ increases.} 
{\color{\change}\begin{lem}\label{4}
	Consider $\Omega \subset [0,1]^{2}$ as a finite set of points and assume that $m+n \le 2$. Then, we have
	\begin{align}
	\mathds{P}\left[ \underset{\bm{r}\in \Omega }{\mathop{\max }}\,{{\left\| \bm{I}_{1}^{(m,n)}(\bm{r}) \right\|}_{2}}\ge \epsilon  \right]\le \delta +\mathds{P}\left[ {{{ \overline{\zeta}}}_{1/4}} \right],~ \forall \delta,\epsilon>0
	\end{align}
provided $	L\ge \tfrac{{{c}_{2}}}{\epsilon^2}\log\Big(\tfrac{12s|\Omega|}{\delta}\Big)(\tfrac{1}{R}\log(\tfrac{2|\Omega|}{\delta})+1)$.
\end{lem}}
Proof. See Appendix \ref{proof.of.lemma2}.
\begin{lem}\label{lem5}
	Let $\Omega \subset [0,1]^{2}$ be a finite set of grid points and $m+n \le 2$. For all $\xi,\epsilon,\delta>0$, with 
	{\color{\change}$	\xi\le\tfrac{c_3\epsilon }{\sqrt{(\tfrac{1}{R}\log(\tfrac{|\Omega|}{\delta})+1)}},$}
	where $c_3 \le \tfrac{1}{4}$, it holds that $\mathds{P}\left[ \underset{\bm{r}\in \Omega }{\mathop{\max }}\,{{\left\| \bm{I}_{2}^{(m,n)} \right\|}_{2}}\ge \epsilon \left| {{\zeta }_{\xi }} \right. \right]\le \delta$.
\end{lem}
Proof. See Appendix \ref{proof.of.lemma3}.

Now we can complete the proof of Lemma \ref{3} by writing
\begin{align}
&\mathds{P}\left[ \underset{\bm{r}\in \Omega }{\mathop{\text{max}}}\,\tfrac{\text{1}}{\kappa^{m+n}}{{\left\| {{\bm{q}}^{(m,n)}}(\bm{r})-{{{ \overline{\bm{q}}}}^{(m,n)}}(\bm{r}) \right\|}_{2}}\ge 2\epsilon  \right]=\mathds{P}\left[ \underset{\bm{r}\in \Omega }{\mathop{\text{max}}}\,\tfrac{\text{1}}{\kappa^{m+n}}{{\left\| \bm{I}_{\text{1}}^{(m,n)}(\bm{r})+\bm{I}_{2}^{(m,n)} \right\|}_{2}}\ge 2\epsilon  \right]\nonumber\\
&\le\mathds{P}\left[ \underset{\bm{r}\in \Omega }{\mathop{\text{max}}}\,\tfrac{\text{1}}{\kappa^{m+n}}{{\left\| \bm{I}_{\text{1}}^{(m,n)}(\bm{r}) \right\|}_{2}}\ge \epsilon  \right]+\mathds{P}\left[ {{\overline{\zeta }}_{\xi }} \right]+\mathds{P}\left[ \underset{\bm{r}\in \Omega }{\mathop{\text{max}}}\,\tfrac{\text{1}}{\kappa^{m+n}}{{\left\| \bm{I}_{\text{2}}^{(m,n)}(\bm{r}) \right\|}_{2}}\ge \epsilon \left| {{\zeta }_{\xi }} \right. \right]\le 4\delta,
\end{align}
where we used the union bound and Lemmas \ref{lem5}, \ref{4}. By choosing
{\color{\change} $\xi=\epsilon c_3(\tfrac{1}{R}\log(\tfrac{|\Omega|}{\delta})+1)^{\tfrac{-1}{2}} $}, the condition in Lemma \ref{2} becomes {\color{\change}$L\ge s(\tfrac{c_1}{\epsilon c_3^{2}})\log^2\tfrac{18s^2}{\delta}(1+\tfrac{1}{R}\log(\tfrac{|\Omega|}{\delta}))$ 
}where $c:=\tfrac{c_1}{c_3^{2}}$. 
\subsubsection{Showing that \texorpdfstring{$\bm{q}(\bm{r})  $}{TEXT} and \texorpdfstring{$\overline{\bm{q}}(\bm{r}) $}{TEXT} are close for all \texorpdfstring{$\bm{r} $}{TEXT} }
In this part, benefiting Lemma \ref{3}, we want to show that ${\bm{q}}^{(m,n)}(\bm{r})$ is close to $\overline{\bm{q}}^{(m,n)}(\bm{r})$ for all  $\bm{r}\in[0,1]^{2}$ with high probability which is given in the following lemma.
	\begin{lem}\label{lem.nearness_all_r} Let $\epsilon,\delta>0$. It holds that
		\begin{align}\label{dif_q(r)}
		\underset{\bm{r}\in[0,1]^{2},(m,n):m+n\le2}{\max}\tfrac{1}{\kappa^{m+n}}\|\bm{q}^{(m,n)}(\bm{r})- \overline{\bm{q}}^{(m,n)}(\bm{r})\|_2\le\epsilon
		\end{align}
		with probability at least $1-\delta$ provided that
		{\color{\change}
		\begin{align}
		L \ge cs \max\{\log^2{\tfrac{12sL^5R}{\epsilon^2\delta}}
		(1+\tfrac{1}{R}\log(\tfrac{L^5R}{\epsilon^2\delta})),\log^2\tfrac{18s^2}{\delta}(1+\tfrac{1}{R}\log(\tfrac{L^5R}{\epsilon^2\delta}))\}
		\end{align}}
	\end{lem}
	To prove, we first choose a set of sufficiently fine points in $\Omega$ such that
	\begin{align}
	\max_{\bm{r}\in [0,1]^2}\min_{\bm{r}_g\in \Omega}\|\bm{r}-\bm{r}_g\|_{\infty}\le \tfrac{\epsilon}{3\tilde{c}L^{\tfrac{5}{2}}R^{\tfrac{1}{2}}}
	\end{align}
	with cardinality
	\begin{align}
	|\Omega|=\Big(\tfrac{3\tilde{c}L^{\tfrac{5}{2}}R^{\tfrac{1}{2}}}{\epsilon}\Big)^{2}=\tfrac{c'L^{5}R}{\epsilon^{2}}.
	\end{align}
	By using the union bound over all six pairs $(m,n)$ obeying $m+n \le 2$ and following Lemma \ref{3}, we find that
	
	\begin{align}\label{dif_q(rg)}
	&\mathds{P}\Big\{\underbrace{\underset{\bm{r}_g\in \Omega,m+n\le 2}{\max}\tfrac{1}{\kappa^{m+n}}\|\bm{q}^{(m,n)}(\bm{r}_g)- \overline{\bm{q}}^{(m,n)}(\bm{r}_g)\|_2\le\tfrac{\epsilon}{3}}_{\mathcal{E}_1}\Big\}\ge 1-\frac{\delta}{2}.
	\end{align}
	To prove that the same result holds for all $\bm{r}\in[0,1]^2$, it is necessary to show that 
	\begin{align}\label{q-event}
	\mathds{P}\Big\{ \underbrace{\underset{\bm{r}\in [0,1]^{2},m+n\le 2}{\max}\tfrac{1}{\kappa^{m+n}}\|\bm{q}^{(m,n)}(\bm{r})\|_2\le\tfrac{\tilde{c}}{2}{L}^{\tfrac{3}{2}} }_{\mathcal{E}_2}\Big\}\ge 1-\frac{\delta}{2},
	\end{align}
	which is proved in Appendix \ref{proof.equation_norm_q}. We also have $\mathds{P}\{\mathcal{E}_1\cap\mathcal{E}_2\}\ge 1-\delta$. Since the event $\mathcal{E}_1\cap \mathcal{E}_2$ implies the event \eqref{dif_q(r)} (see Appendix \ref{proof.whyimply} for the reason), Lemma \ref{lem.nearness_all_r} is concluded.
	\subsubsection{Showing that \texorpdfstring{$\|\bm{q}(\bm{r})\|_2\le 1$}{TEXT} for all \texorpdfstring{$\bm{r} \notin \mathcal{S}$}{TEXT}}\label{proof.offsupoort}
	We begin with defining the following sets:
	\begin{align}
	\Omega_{\rm far}	:= \forall \bm{r}\in [0,1]^{2} :\min_{\bm{r}_j\in \mathcal{S}}\|\bm{r}-\bm{r}_j\|_{\infty}\ge \tfrac{0.2447}{N}\label{eq.far}\\
	\Omega_{\rm close}:= \forall \bm{r}\notin \mathcal{S},\bm{r}_j \in \mathcal{S} :0\le \|\bm{r}-\bm{r}_j\|_{\infty}\le \tfrac{0.2447}{N}\label{eq.near}.
	\end{align}
	The former argument \eqref{eq.far} implies that the points are far from $\bm{r}_j$ while the latter \eqref{eq.near} include points that are close to it. In order to show that $\bm{q}(\bm{r})$ in \eqref{dual-poly} satisfies \eqref{dual}, it is enough to show that $\|\bm{q}(\bm{r})\|_2 \le 1$ for $\forall \bm{r}\in \Omega_{\rm far}$ and $\forall \bm{r}\in \Omega_{\rm close}$. To proceed,	suppose that 
	{\color{\change}
	\begin{align}
	L \ge cs \max\{\log^2{\tfrac{12sL}{\delta}}
(1+\tfrac{1}{R}\log(\tfrac{2L}{\delta})),\log^2\tfrac{18s^2}{\delta}(1+\tfrac{1}{R}\log(\tfrac{L}{\delta}))\}.
	\end{align}
}
	\begin{lem}\label{omega-far}
		$\|\bm{q}(\bm{r})\|_2\le 1, \forall \bm{r}\in \Omega_{\rm far}$	 with high probability.
		
	\end{lem}
	To prove this, take $\epsilon=0.002$ in \eqref{dif_q(r)} to reach
	\begin{align}
	\|\bm{q}^{(m,n)}(\bm{r})- \overline{\bm{q}}^{(m,n)}(\bm{r})\|_2 \le 0.002. 
	\end{align}
	By the triangular inequality, we have 
	\begin{align}
	&\|\bm{q}^{(m,n)}(\bm{r})\|_2\le\|\bm{q}^{(m,n)}(\bm{r})- \overline{\bm{q}}^{(m,n)}(\bm{r})\|_2+\| \overline{\bm{q}}^{(m,n)}(\bm{r})\|_2\le0.9978,
	\end{align}
	which verifies that $\|\bm{q}(\bm{r})\|_2\le 1$ for far $\bm{r}$. For near $\bm{r}$, we state the following lemma.
	\begin{lem}\label{lem.near}
		$\|\bm{q}(\bm{r})\|_2\le 1, \forall \bm{r}\in \Omega_{\rm close}$
		with high probability.
	\end{lem}
	To prove this, assume without loss of generality that $\bm{0}\in \mathcal{S} $ i.e. $\|\bm{r}\|_{\infty}\le\tfrac{0.2447}{N}$. A sufficient condition for Lemma \ref{lem.near} to hold, is to show that the Hessian matrix of $\|\bm{q}(\bm{r})\|_2^{2}$, i.e.,
	\begin{align}
	\tfrac{1}{\kappa^{2}}\bm{H}=\tfrac{1}{\kappa^{2}} \begin{bmatrix}
	\tfrac{\partial^{2}\|\bm{q}(\bm{r})\|_2^{2}}{\partial\tau^{2}}& \tfrac{\partial^2\|\bm{q}(\bm{r})\|_2^{2}}{\partial\tau\partial\nu}\\
	&\\
	\tfrac{\partial^2 \|\bm{q}(\bm{r})\|_2^{2}}{\partial\tau\partial\nu}&\tfrac{\partial^{2}\|\bm{q}(\bm{r})\|_2^{2}}{\partial\nu^{2}}
	\end{bmatrix} 
	\end{align}
	is negative definite for all near $\bm{r}$. For this to hold, we should have
	\begin{align}\label{74}
	\tfrac{1}{\kappa^{2}}{\rm tr}(\bm{H})=\tfrac{\partial^{2}}{\partial\tau^{2}}\tfrac{1}{\kappa^{2}}\|\bm{q}(\bm{r})\|_2^{2}+\tfrac{\partial^{2}}{\partial\nu^{2}}\tfrac{1}{\kappa^{2}}\|\bm{q}(\bm{r})\|_2^{2} < 0
	\end{align}
	and
	\begin{align}\label{determinan}
	&\tfrac{1}{\kappa^{2}}{\rm det}(\bm{H})=\Big(\tfrac{\partial^{2}}{\partial\tau^{2}}\tfrac{1}{\kappa^{2}}\|\bm{q}(\bm{r})\|_2^{2}\Big)\Big(\tfrac{\partial^{2}}{\partial\nu^{2}}\tfrac{1}{\kappa^{2}}\|\bm{q}(\bm{r})\|_2^{2}\Big)-\Big(\tfrac{\partial^2}{\partial\tau\partial\nu}\tfrac{1}{\kappa^{2}}\|\bm{q}(\bm{r})\|_2^{2}\Big)^{2} > 0.
	\end{align}
	To find $\|\bm{q}(\bm{r})\|_2$ and its derivatives, we need to borrow some bounds from \cite[Section, C.2]{candes2014towards} which states that $\forall \bm{r}\in \Omega_{{\rm close}}$ and $N\ge 512$, we have
	\begin{align}
	\overline{G}(\bm{r})\ge 0.8113,\quad 	 | \overline{G}^{(1,0)}(\bm{r})|\le 0.8113,\nonumber\\
	\overline{G}^{(2,0)}(\bm{r})\le -2.097N^{2},\quad 	  | \overline{G}^{(1,1)}(\bm{r})|\le 0.6531N,\nonumber\\
	| \overline{G}^{(2,1)}(\bm{r})|\le 2.669N^{2},\quad 	 | \overline{G}^{(3,0)}(\bm{r})|\le 8.070N^{3}.
	\end{align}
	Introduce
	\begin{align}
	\overline{Z}^{(m',n')}(\bm{r}):=\sum_{\bm{r}_j\in \mathcal{S} \setminus {{0}}}| \overline{G}^{(m',n')}(\bm{r}-\bm{r}_j)|.
	\end{align}
	Again, based on \cite[Table C.1]{candes2014towards}, it holds that
	\begin{align}
	\overline{Z}^{(0,0)}(\bm{r}) \le 6.405\times 10^{-2},\quad \overline{Z}^{(1,0)}(\bm{r}) \le 0.1047N,\nonumber\\
	\overline{Z}^{(2,0)}(\bm{r}) \le0.4019N,\quad \overline{Z}^{(1,1)}(\bm{r})\le 0.1642N^{2},\nonumber\\
	\overline{Z}^{(2,1)}(\bm{r}) \le 0675N^{2},\quad \overline{Z}^{(3,0)}(\bm{r})\le 1.574N^{3},
	\end{align}
	and we also have \cite[Section C.1]{candes2014towards} 
	\begin{align}
	\| \overline{\bm{\alpha}}_j\|_{2}\le \alpha_{\max}=1+5.577\times 10^{-2}\nonumber\\
	\| \overline{\bm{\alpha}}_j\|_{2} \ge  \alpha_{\min}=1-5.577\times 10^{-2}\nonumber\\
	\| \overline{\bm{\beta}}_j\|_{2}\le \beta_{\max}=\tfrac{2.93}{N}\times 10^{-2}\nonumber\\
	\| \overline{\bm{\gamma}}_j\|_{2}\le \gamma_{\max}=\tfrac{2.93}{N}\times 10^{-2}.
	\end{align}
	We use the aforementioned formulas to obtain the bounds
	\begin{align}
	&\| \overline{\bm{q}}(\bm{r})\|_2=\|\sum_{j=1}^{s} \overline{G}^{(0,0)}(\bm{r}-\bm{r}_j) \overline{\bm{\alpha}}_j +  \overline{G}^{(1,0)} (\bm{r}-\bm{r}_j) \overline{\bm{\beta}}_j
	\nonumber\\
	&+ \overline{G}^{(0,1)}(\bm{r}-\bm{r}_j) \overline{\bm{\gamma}}_j\|_2\le \alpha_{\max}\Big(| \overline{G}^{(0,0)}(\bm{r})|+ \overline{Z}^{(0,0)}(\bm{r})\Big)\nonumber\\
	&+ 2\beta_{\max}\Big(| \overline{G}^{(1,0)}(\bm{r})|+ \overline{Z}^{(1,0)}(\bm{r})\Big)\le 1.295 + \tfrac{0.0475}{N}.
	\end{align}
	For the derivatives of $\bm{q}$, we have:
	\begin{align}\label{q(1,0)}
	&\| \overline{\bm{q}}^{(1,0)}\|_2\le \alpha_{\max}\Big(| \overline{G}^{(1,0)}(\bm{r})|+ \overline{Z}^{(1,0)}(\bm{r})\Big)\nonumber\\
	&+ \beta_{\max}\Big(| \overline{G}^{(2,0)}(\bm{r})|+ \overline{Z}^{(2,0)}(\bm{r})\Big)+\gamma_{\max}\Big(| \overline{G}^{(1,1)}(\bm{r})|\nonumber\\
	&+ \overline{Z}^{(1,1)}(\bm{r})\Big)\le 0.08874 +0.2148N.	
	\end{align} 
	Other derivatives can be obtained using similar steps as follows:
	\begin{align}\label{bound}
	&\| \overline{\bm{q}}^{(1,1)}\|_2\le 0.846N +0.213N^{2},\nonumber\\
	&\| \overline{\bm{q}}^{(2,0)}\|_2\le 0.5025N +3.8845N^{2}.
	\end{align}
	Now, we proceed \eqref{74} by writing
	\begin{align}\label{80}
	&\tfrac{\partial^{2}}{\partial\tau^{2}}\|\tfrac{1}{\kappa}\bm{q}(\bm{r})\|^{2}_2=\tfrac{\partial}{\partial\tau}\tfrac{1}{\kappa^{2}}\langle \bm{q}^{(1,0)}(\bm{r}),\bm{q}(\bm{r})\rangle=2\| \tfrac{1}{\kappa}\bm{q}^{(1,0)}(\bm{r})\|^{2}_{2}+ \tfrac{2}{{\kappa}^{2}}{\rm Re}\Big[\Big(\bm{q}^{(2,0)}(\bm{r})\Big)^{\rm H}\bm{q}(\bm{r})\Big],
	\end{align}
	where the first term can be bounded as 
	\begin{align}\label{81}
	&\| \tfrac{1}{\kappa}\bm{q}^{(1,0)}(\bm{r})\|^{2}_{2}\le \| \tfrac{1}{\kappa}\Big(\bm{q}^{(1,0)}(\bm{r})- \overline{\bm{q}}^{(1,0)}(\bm{r})\Big)\|^{2}_{2}+\| \tfrac{1}{\kappa} \overline{\bm{q}}^{(1,0)}(\bm{r})\|^{2}_{2}\le \epsilon^{2}+0.0141.
	\end{align}
	The first inequality above comes from the triangular inequality while the last one is based on Lemma \ref{lem.nearness_all_r}, \ref{q(1,0)} and the fact that ${\kappa}^{2} \ge \tfrac{\pi^{2}}{3}N^{2}$. The second term in \eqref{80} can be bounded by:
	\begin{align}\label{82}
	&\tfrac{1}{{\kappa}^{2}}{\rm Re}\Big[\Big(\bm{q}^{(2,0)}(\bm{r})\Big)^{\rm H}\bm{q}(\bm{r})\Big]={\rm Re}\Big[\tfrac{2}{{\kappa}^{2}}\Big(\bm{q}^{(2,0)}(\bm{r})\nonumber\\
	&- \overline{\bm{q}}^{(2,0)}(\bm{r})+ \overline{\bm{q}}^{(2,0)}(\bm{r})\Big)^{\rm H}(\bm{q}(\bm{r})- \overline{\bm{q}}(\bm{r})+ \overline{\bm{q}}(\bm{r}))\Big]=\nonumber\\
	&{\rm Re}\Big[\tfrac{1}{{\kappa}^{2}}\Big(\bm{q}^{(2,0)}(\bm{r})- \overline{\bm{q}}^{(2,0)}(\bm{r})\Big)^{\rm H}(\bm{q}(\bm{r})- \overline{\bm{q}}(\bm{r}))\Big]\nonumber\\
	&+{\rm Re}\Big[\tfrac{1}{{\kappa}^{2}}\Big( \overline{\bm{q}}^{(2,0)}(\bm{r})\Big)^{\rm H} \overline{\bm{q}}(\bm{r})\Big]\nonumber\\
	&+{\rm Re}\Big[\tfrac{1}{{\kappa}^{2}}\Big(\bm{q}^{(2,0)}(\bm{r})- \overline{\bm{q}}^{(2,0)}(\bm{r})\Big)^{\rm H} \overline{\bm{q}}(\bm{r})\Big]\nonumber\\
	&+{\rm Re}\Big[\tfrac{1}{{\kappa}^{2}}\Big( \overline{\bm{q}}^{(2,0)}(\bm{r})\Big)^{\rm H}(\bm{q}(\bm{r})- \overline{\bm{q}}(\bm{r}))\Big]\nonumber\\
	&\le \epsilon^{2}-0.307+1.129\epsilon +1.181\epsilon\nonumber\\
	&\le \epsilon^{2}+ 2.31\epsilon - 0.307.
	\end{align}
	Substituting \eqref{82} and \eqref{81} into \eqref{80}, yields to
	\begin{align}\label{86}
	\tfrac{1}{{\kappa}^{2}} {\rm tr}(\bm{H})\le 8\epsilon^{2}+ 9.24\epsilon - 1.1712.
	\end{align}
	It is straightforward to verify that the above term is negative by setting $\epsilon\le 0.1$. Next, we prove \eqref{determinan}. 
	The second term in \eqref{determinan} can be written as
	\begin{align}\label{87}
	&\tfrac{\partial}{\partial\tau \partial\nu}\|\tfrac{1}{\kappa} \bm{q}(\bm{r})\|_2^{2}=\tfrac{2}{{\kappa}^{2}}\langle \bm{q}^{(1,0)}(\bm{r}),\bm{q}^{(0,1)}(\bm{r}) \rangle+\tfrac{2}{{\kappa}^{2}}\langle \bm{q}^{(1,1)}(\bm{r}),\bm{q}(\bm{r}) \rangle.
	\end{align}
	The upper-bound of the first term in \eqref{determinan} is obtained by
	\begin{align}\label{88}
	&\tfrac{1}{{\kappa}^{2}}\langle \bm{q}^{(1,0)}(\bm{r}),\bm{q}^{(0,1)}(\bm{r}) \rangle={\rm Re} \Big[\tfrac{1}{{\kappa}^{2}}\Big(\bm{q}^{(1,0)}(\bm{r})- \overline{\bm{q}}^{(1,0)}(\bm{r})\Big)^{\rm H}\nonumber\\
	&\Big(\bm{q}^{(0,1)}(\bm{r})- \overline{\bm{q}}^{(0,1)}(\bm{r})\Big) \Big]+{\rm Re} \Big[\tfrac{1}{{\kappa}^{2}}\Big( \overline{\bm{q}}^{(1,0)}(\bm{r})\Big)^{\rm H} \overline{\bm{q}}^{(0,1)}(\bm{r}) \Big]\nonumber\\
	&+{\rm Re}\Big[\tfrac{1}{{\kappa}^{2}}\Big(\bm{q}^{(1,0)}(\bm{r})- \overline{\bm{q}}^{(1,0)}(\bm{r})\Big)^{\rm H} \overline{\bm{q}}^{(0,1)}(\bm{r}) \Big]\nonumber\\
	&+{\rm Re}\Big[\tfrac{1}{{\kappa}^{2}}\Big( \overline{\bm{q}}^{(1,0)}(\bm{r})\Big)^{\rm H}\Big(\bm{q}^{(0,1)}(\bm{r})- \overline{\bm{q}}^{(0,1)}(\bm{r})\Big) \Big]\nonumber\\
	&\le \epsilon^{2} + 0.238\epsilon +0.0736,
	\end{align}
	where the last inequality follows from \eqref{dif_q(r)},\eqref{bound} and the fact that ${\kappa}^{2} \ge \tfrac{\pi^{2}}{3}N^{2}$. By using similar steps as in \eqref{88}, we reach
	\begin{align}\label{89}
	\tfrac{1}{\kappa^{2}}\langle \bm{q}^{(1,1)}(\bm{r}),\bm{q}(\bm{r})\rangle \le \epsilon^{2} +0.195\epsilon +0.0736.
	\end{align}
	Substituting \eqref{88} and \eqref{89} into \eqref{87} yields to
	\begin{align}\label{moshtagh2bodi}
	\tfrac{\partial^2}{\partial\tau \partial\nu}\|\tfrac{1}{{\kappa}}\bm{q}(\bm{r})\|_2^{2}\le 4\epsilon^{2}+ 2.865\epsilon +0.175.
	\end{align}
	By using the bound obtained for \eqref{moshtagh2bodi} and \eqref{80} and setting $\epsilon=0.05$,~\eqref{determinan} is satisfied.
	Finally, based on Lemmas \ref{omega-far} and \ref{lem.near}, we can show that $\|\bm{q}(\bm{r})\|_2 \le 1, \forall \bm{r}\in [0,1]^{2}\ \setminus \mathcal{S}$.
	\subsection{Proof of Lemma \ref{4}}\label{proof.of.lemma2}
	
	Set $\epsilon=2.5ab$ and $\Delta \bm{w}^{(m,n)}:=\bm{w}^{(m,n)}(\bm{r})-\overline{\bm{w}}^{(m,n)}(\bm{r})$. For all $a,b\ge 0$ we have
	\begin{align}\label{eq.rel4}
	&\mathds{P}[\underset{\mathbf{ r} \in \Omega}{\max}\| \bm{I}_1^{(m,n)}(\bm{r})\|_2 \ge 2.5ab]=\mathds{P}\Big[\underset{\bm{r}\in\Omega}{\max}\|(\Delta \bm{w}^{(m,n)})^{\text{H}} \mathbf{L} \bm{\Phi}\|_2 \nonumber \\
	&\ge 2.5ab\Big] \le \mathds{P}\Big[\underset{\bm{r} \in \Omega }{\bigcup}\Big\{\|(\Delta \bm{w}^{(m,n)})^{\text{H}} \mathbf{L} \bm{\Phi}\|_2 \ge \| \mathbf{L}^{\text{H}}(\Delta \bm{w}^{(m,n)}) \|_2b\Big\}\nonumber\\
	& \cup \Big\{\|\mathbf{L}^{\text{H}}(\Delta \bm{w}^{(m,n)})\|_2\ge 2.5a\big\} \Big]\le \mathds{P}\Big[\underset{r\in\Omega}{\bigcup}\Big\{\|(\Delta \bm{w}^{(m,n)})^{\text{H}} \mathbf{L} \bm{\Phi}\|_2 \nonumber\\
	&\ge \| \mathbf{L}^{\text{H}}\Delta \bm{w}^{(m,n)} \|_2b\Big\}\cup\Big\{\|\Delta \bm{w}^{(m,n)}\|_2\ge a \Big\}\cup \Big\{\|\mathbf{L}\|\ge2.5\Big\}\Big]\nonumber\\
	&\le \mathds{P}[\|\mathbf{L}\|\ge2.5]+\sum_{\bm{r} \in \Omega}\Big(\mathds{P}\Big[\|(\Delta \bm{w}^{(m,n)})^{\text{H}} \mathbf{L} \bm{\Phi}\|_2\nonumber\\
	&\stackrel{(\RN{1})}{\ge} \| \mathbf{L}^{\text{H}}(\Delta \bm{w}^{(m,n)}) \|_2b\Big]+\mathds{P}\Big[\|\Delta \bm{w}^{(m,n)}\|_2\ge a\Big]\Big)\nonumber\\
	&\le \mathds{P}[ \overline{\zeta}_{\tfrac{1}{4}}]+(R+1)|\Omega|e^{-\tfrac{b^2}{8}}+\mathds{P}\Big[\|\Delta \bm{w}^{(m,n)}\|_2\ge a\Big]\nonumber\\
	&\le \mathds{P}[ \overline{\zeta}_{\tfrac{1}{4}}]+\tfrac{\delta}{2}+\mathds{P}\Big[\|\Delta \bm{w}^{(m,n)}\|_2\ge a \Big],
	\end{align}
	where $(\RN{1})$ comes from the generalized Bersntein inequality given below.
	\begin{lem}
		{\color{\chang}\cite[Lemma 4]{yang2018sample} Let the rows of $\bm{\Phi} \in \mathbb{C}^{s\times R}$ be sampled
		independently on the complex hyper-sphere $\mathbb{S}^{R-1}$ with zero mean. Then, for all $\mathbf{L}^{\text{H}}\Delta \bm{w}^{(m,n)} \in \mathbb{C}^{s},~\mathbf{L}^{\text{H}}\Delta \bm{w}^{(m,n)}\ne \bm{0}$ and $b\ge0$, 
		\begin{align*}
		\mathds{P}(\|(\Delta \bm{w}^{(m,n)})^{\text{H}} \mathbf{L} \bm{\Phi}\|_2\ge \| \mathbf{L}^{\text{H}}\Delta \bm{w}^{(m,n)} \|_2b)\le  e^{-R({b^{2}}-\log(b^{2})-1)}.
		\end{align*}
	}
	{\color{\change}By using $|\Omega|e^{-R({b^{2}}-\log(b^{2})-1)\le\tfrac{\delta}{2}}$, we obtain $b^2=(\tfrac{1}{R}\log(\tfrac{2|\Omega|}{\delta})+1)$ .}
	\end{lem}
{\color{\change}By using \cite[Section 8.3.1]{heckel2016super} and choosing $\mathds{P}[\|\Delta \bm{w}^{(m,n)}\|_2 \ge a]\le \tfrac{\delta}{2}$ and $a=\sqrt{\tfrac{3s}{L}}12^{\tfrac{3}{2}}c_1(\tfrac{c^{2}_2}{c})$, we proceed \eqref{eq.rel4} as follows:
	\begin{align}
	&\mathds{P}\Big[\underset{\bm{r}\in \Omega}{\text{max}}\|\bm{I}_1^{(m,n)}(\bm{r})\|_2\ge 360c_1(\tfrac{c_2^{2}}{\sqrt{c}}) \sqrt{\tfrac{s}{L}}\nonumber\\
	&\log\Big(\tfrac{12s|\Omega|}{\delta}\Big)\sqrt{(\tfrac{1}{R}\log(\tfrac{2|\Omega|}{\delta})+1)}\Big]\le \delta+\mathds{P}(\overline{\zeta}_{\frac{1}{4}})=2\delta.
	\end{align}
}

	\subsection{Proof of Lemma \ref{lem5}:}\label{proof.of.lemma3}
{\color{\change}	By the union bound, we have:
	\begin{align}
	&\mathds{P}\Big[\underset{\bm{r}\in\Omega}{\text{max}}\|\bm{I}_2^{(m,n)}(\bm{r})\|_2\ge \epsilon~ \big|~ \zeta_\xi\Big]\le\sum_{\bm{r}\in\Omega}\mathds{P}\Big[\|( \overline{\bm{w}}^{(m,n)}(\bm{r}))^{\text{H}}(\mathbf{L}- \overline{\mathbf{L}})\bm{\Phi} \|_2\ge \epsilon~ \big|~ \zeta_\xi\Big]\nonumber\\
	&\le\sum_{\bm{r}\in\Omega}\mathds{P}\Big[\|( \overline{\bm{w}}^{(m,n)}(\bm{r}))^{\text{H}}(\mathbf{L}- \overline{\mathbf{L}})\bm{\Phi}\|_2\nonumber\\& \ge \|(\mathbf{L}- \overline{\mathbf{L}})^{\text{H}} \overline{\bm{w}}^{(m,n)}(\bm{r})\|_2\tfrac{\epsilon}{(c_2\xi)} \Big]\le |\Omega|e^{-R({(\tfrac{\epsilon}{(c_2\xi)})^{2}}-\log((\tfrac{\epsilon}{(c_2\xi)})^{2})-1)}\le\delta.
	\end{align}
}

\subsubsection{Proof of  \eqref{q-event}}\label{proof.equation_norm_q}
We first find an upper-bound for $\|\bm{q}^{(m,n)}(\bm{r})\|_2$ as follows:
\begin{align}\label{bound_q}
&\tfrac{1}{\kappa^{m+n}}\|\bm{q}^{(m,n)}(\bm{r})\|_2=\|(\bm{w}^{(m,n)})^{\rm H}\mathbf{L}\bm{\Phi}\|_2\le\|\mathbf{L}\| \|\bm{\Phi}\|_F\|\bm{w}^{(m,n)}\|_2\le\|\mathbf{L}\| \sqrt{s} \sqrt{3s} \|\bm{w}^{(m,n)}\|_\infty \nonumber\\
& \le\|\mathbf{L}\|s \sqrt{3}\underset{j,({m}',{n}')\in \{(0,0,),(1,0),(0,1)\}}{\max} \tfrac{|G_{({m}',{n}')}^{(m,n)}(\bm{r},{{\bm{r}}_{j}})|}{\kappa^{m+{m}'+{n}'+n}},
\end{align}
for all $\bm{r}$ and all $\bm{r}_j$ we have \cite[Equation 8.66]{heckel2016super}:
\begin{align}\label{bound_G}
\tfrac{|G_{({m}',{n}')}^{(m,n)}(\bm{r},{{\bm{r}}_{j}})|}{\kappa^{m+{m}'+{n}'+n}}\le c_1 12^{\tfrac{3}{2}}\sqrt{L}\|\bm{x}\|^{2}_2.
\end{align} 
By replacing \eqref{bound_G} into \eqref{bound_q} and using $s\le L$, we have
$\tfrac{\|\bm{q}^{(m,n)}(\bm{r})\|_2}{\kappa^{m+n}}\le 72 c_1L^{\tfrac{3}{2}}\|\mathbf{L}\|\|\bm{x}\|^{2}_2$. Thus, by taking $\tfrac{\tilde{c}}{2}=(2.5)\cdotp(3)\cdotp(72)c_1$,
\begin{align}
&\mathds{P}\Big[\underset{\bm{r}\in[0,1]^{2},m+n\le2}{\max}\|\bm{q}^{(m,n)}(\bm{r})\|_2\ge \tfrac{\tilde{c}}{2} L^{\tfrac{3}{2}}\Big]\le \mathds{P}[\|\mathbf{L}\|\|\bm{x}\|^{2}_2\ge\nonumber\\
&(2.5)\cdotp(3)]\le \mathds{P}[\|\mathbf{L}\|\ge2.5]+\mathds{P}[\|\bm{x}\|^{2}_2\ge 3]\le \tfrac{\delta}{2},
\end{align}
where we used the fact that $\mathds{P}[\|\bm{x}\|^{2}_2\ge 3] \le \tfrac{\delta}{4}$ \cite[Equation 8.69]{heckel2016super}. The last inequality follows from $\mathds{P}[\|\mathbf{L}\|\ge 2.5]\le \mathds{P}[ \overline{\zeta}_{\tfrac{1}{4}}]\le\tfrac{\delta}{4}$ (from Lemma \ref{lem.L_bounds}).
\subsection{The reason that $\mathcal{E}_1$, $\mathcal{E}_2$ imply \eqref{dif_q(r)}}\label{proof.whyimply}
Let $\bm{r}_g$ be the closest points in $\Omega$ to $\bm{r}$ with respect to $\ell_{\infty}$-measure. By the triangle inequality:
\begin{align}\label{64}
& \tfrac{1}{\kappa^{m+n}}\|\bm{q}^{(m,n)}(\bm{r})- \overline{\bm{q}}^{(m,n)}(\bm{r})\|_2\le \tfrac{1}{\kappa^{m+n}}\Big[ \|\bm{q}^{(m,n)}(\bm{r})-{\bm{q}}^{(m,n)}(\bm{r}_g)\|_2 + \nonumber\\
&\|\bm{q}^{(m,n)}(\bm{r}_g)- \overline{\bm{q}}^{(m,n)}(\bm{r}_g)\|_2 +\| \overline{\bm{q}}^{(m,n)}(\bm{r}_g)- \overline{\bm{q}}^{(m,n)}(\bm{r})\|_2\Big].
\end{align}
Next, we obtain upper-bounds for the composing terms of the above relation, separately. For the first term, we have:
\begin{align}\label{65}
&\|\bm{q}^{(m,n)}(\bm{r})-{\bm{q}}^{(m,n)}(\bm{r}_g)\|_2\le \sqrt{R}~\underset{i}{\max}|\bm{q}^{(m,n)}(\bm{r})-{\bm{q}}^{(m,n)}(\bm{r}_g)|_i.
\end{align}
We proceed \eqref{65} by writing
\begin{align}\label{bernestein}
&|\bm{q}^{(m,n)}(\bm{r})-{\bm{q}}^{(m,n)}(\bm{r}_g)|_i=|\bm{q}^{(m,n)}(\tau,\nu)-\bm{q}^{(m,n)}(\tau,\nu_g)+\bm{q}^{(m,n)}(\tau,\nu_g)\nonumber\\
&-\bm{q}^{(m,n)}(\tau_g,\nu_g)|_i\le|\bm{q}^{(m,n)}(\tau,\nu)-\bm{q}^{(m,n)}(\tau,\nu_g)|_i+|\bm{q}^{(m,n)}(\tau,\nu_g)-\bm{q}^{(m,n)}(\tau_g,\nu_g)|_i\nonumber\\
&\le|\nu-\nu_g|\underset{z}{\sup}|\bm{q}^{(m,n+1)}(\tau,z)|_i+|\tau-\tau_g|\underset{z}{\sup}|\bm{q}^{(m+1,n)}(z,\nu_g)|_i\nonumber\\
&\le |\nu-\nu_g|2\pi N \underset{z}{\sup}\|\bm{q}^{(m,n)}(\tau,z)\|_2 + |\tau-\tau_g|2\pi N \underset{z}{\sup}\|\bm{q}^{(m,n)}(z,\tau)\|_2,
\end{align}
where in the last step, we used Bernstein's polynomial inequality \cite[Corollary 8]{harris1996bernstein}. Substituting \eqref{q-event} into \eqref{bernestein}, we reach
\begin{align}\label{ber2}
&\tfrac{1}{\kappa^{m+n}}\|\bm{q}^{(m,n)}(\bm{r})-\bm{q}^{(m,n)}(\bm{r}_g)\|_2 \nonumber\\
&\le \tfrac{\tilde{c}}{2}L^{\tfrac{5}{2}}R^{\tfrac{1}{2}}(|\tau -\tau_g|+|\nu-\nu_g|)\nonumber\\
&\le \tilde{c}L^{\tfrac{5}{2}}R^{\tfrac{1}{2}}\|\bm{r}-\bm{r}_g\|_{\infty}\le \tfrac{\epsilon}{3}.
\end{align}
Similarly, we have
\begin{align}\label{bound_q_ber}
\tfrac{1}{\kappa^{m+n}}\| \overline{\bm{q}}^{(m,n)}(\bm{r}_g)- \overline{\bm{q}}^{(m,n)}(\bm{r})\|_2\le\tfrac{\epsilon}{3}.
\end{align}

Substituting \eqref{dif_q(rg)}, \eqref{ber2}, \eqref{bound_q_ber} into \eqref{64} leads to ${\tfrac{1}{\kappa^{m+n}}\|\bm{q}^{(m,n)}(\bm{r})- \overline{\bm{q}}^{(m,n)}(\bm{r})\|_2\le \epsilon}$ for all $(m,n):m+n\le 2$ and for all $\bm{r}\in [0,1]^{2}$.

\section*{Acknowledgements}
The authors would like to thank Reinhard Heckel for helpful comments and suggestions.
\section*{References}

\bibliography{mypaperbibe1}

\end{document}